\documentclass{article}
\usepackage{amsmath,amsthm}

\newtheorem{theorem}{Theorem}[section]
\newtheorem{lemma}[theorem]{Lemma}
\newtheorem{definition}[theorem]{Definition}
\newtheorem{corol}[theorem]{Corollary}

\newenvironment{remark}%
  {\par\medbreak\refstepcounter{theorem}%
    \noindent\textbf{Remark~\thetheorem. }}%
  {\par\medskip}

\newcommand{\vz}[1]{\ensuremath{\mathbb{#1}}}

\newcommand{\R}{{\vz R}}
\newcommand{\N}{{\vz N}}
\DeclareMathOperator{\supp}{supp}
\def\pref#1{(\ref{#1})}

\long\def\drop#1{}
\def\longrightharpoonup{\relbar\joinrel\rightharpoonup}
\let\weakto\longrightharpoonup
\def\weakstarto{\overset{*}\longrightharpoonup}
\let\e\varepsilon
\let\epsilon\varepsilon
\def\HNo{\mathcal H^{N-1}}
\let\ds\displaystyle
\def\Xint#1{\mathchoice
   {\XXint\displaystyle\textstyle{#1}}%
   {\XXint\textstyle\scriptstyle{#1}}%
   {\XXint\scriptstyle\scriptscriptstyle{#1}}%
   {\XXint\scriptscriptstyle\scriptscriptstyle{#1}}%
   \!\int}
\def\XXint#1#2#3{{\setbox0=\hbox{$#1{#2#3}{\int}$}
     \vcenter{\hbox{$#2#3$}}\kern-.5\wd0}}
\def\dashint{\Xint-}

\usepackage{a4wide}
\usepackage{a4}
\usepackage{latexsym}
\usepackage{amssymb}
\usepackage{amsfonts}
\usepackage{bm}
\usepackage{epsfig}
\usepackage{bbm}
\usepackage{amsmath}
\usepackage{graphicx}

\usepackage{psfrag}
\usepackage{color}
\usepackage{multicol}
\usepackage{subfig}

\begin{document}
\title{Copolymer-homopolymer blends: global energy minimisation and global
energy bounds}
\author{Yves van Gennip \and Mark A. Peletier}
\date{\today}
\maketitle
\begin{abstract}
We study a variational model for a diblock copolymer-homopolymer blend. The energy functional is a sharp-interface limit of a generalisation of the Ohta-Kawasaki energy. In one dimension, on the real line and on the torus, we prove existence of minimisers of this functional and we describe in complete detail the structure and energy of stationary points. Furthermore we characterise the conditions under which the minimisers may be non-unique.

In higher dimensions we construct lower and  upper bounds on the energy of minimisers, and explicitly compute the energy of spherically symmetric configurations.\\[2\jot]
\textbf{Keywords:} block copolymers, copolymer-homopolymer blends, pattern formation, variational model, partial localisation, lipid bilayers\\[2\jot]
\textit{Mathematics Subject Classification (2000):} 49N99, 82D60
\end{abstract}

\section{Introduction}
\subsection{Micro-phase separation}

In this paper we study the functional
\begin{equation}\label{eq:functional}
F_1(u, v) = \left\{ \begin{array}{ll} \displaystyle
c_0 \int_{\R^N} |\nabla (u + v)| + c_u \int_{\R^N} |\nabla u| + c_v \int_
{\R^N} |\nabla v|\hspace{0.25cm} + \|u - v\|_{H^{-1}(\R^N)}^2
  &\hspace{-.35cm} \mbox{ if $(u, v) \in K_1$,}\vspace{0.25cm}\\
\infty &\hspace{-.35cm} \mbox{ otherwise,} \end{array} \right.
\end{equation}
where the coefficients $c_i$ are nonnegative (not all equal to zero) and \footnote{Where we do not explicitly specify the integration measure, we use the Lebesgue measure.}
\[
K_1 := \left\{ (u, v) \in \left(\text{BV}(\R^N)\right)^2 :
   u(x), v(x) \in \{0, 1\} \text{ a.e., and } uv = 0 \text{ a.e., and }
   \int_{\R^N} u = \int_{\R^N} v \; \right\}.
\]

\medskip

Under the additional constraint $u+v\equiv 1$, this functional is the sharp-interface limit of a well-studied variational model for melts
of diblock copolymers~\cite{Choksi01,ChoksiRen03,ChoksiSternberg06,FifeHilhorst01,Muratov02,RenWei00,RenWei02,RenWei03a,RenWei03b,RenWei05,RenWei06a,RenWei06b}. This underlying diffuse interface model is also closely related to the functional studied in \cite{Mueller93}. Such polymers consist of two parts, labelled the U and V parts, whose volume fractions are represented by the variables $u$ and $v$. The U and V parts
of the polymers repel each other, and this repulsion leads to \emph{micro-phase separation}:
phase separation at a length scale comparable to the length of a single
molecule. The case studied here is known as the \emph{strong segregation limit}, \cite{BatesFredrickson99},
in which strong repulsion causes strong demixing of the constituents---hence the
restriction of $K_1$ to characteristic functions. The modeling assumption here is that stationary points of $F_1$ under constrained (i.e. fixed) mass $\int_{\R^N} u$, in particular minimisers, represent the structures formed by the polymers.

Although the various simplifications leading to $F_1$ have obscured the
connection between this functional and single molecules, the character of the
various terms is still recognisable. The interfacial penalisation terms, i.e. the first three terms, are what
remains of the repulsion in the strong segregation limit, and these terms favour
large-scale demixing. The last term $\|u-v\|_{H^{-1}}$, on the other hand,
penalises such large-scale separation and arises from the chemical bond between
the U and V parts of the polymer molecules.

\begin{figure}[ht]
\subfloat[Diblock copolymer][$n$-block structure on the unit torus $\vz{T}_1$, with $u+v\equiv 1$]
{
    \psfrag{a}{\Huge{\boldmath $\overbrace{\hspace{6,825cm}}$ \unboldmath}}
    \psfrag{b}{$0$}
    \psfrag{c}{$1$}
    \psfrag{d}{\Large{\boldmath $2n$ \unboldmath}}
    \psfrag{1}{$1$}
    \includegraphics[width=70mm]{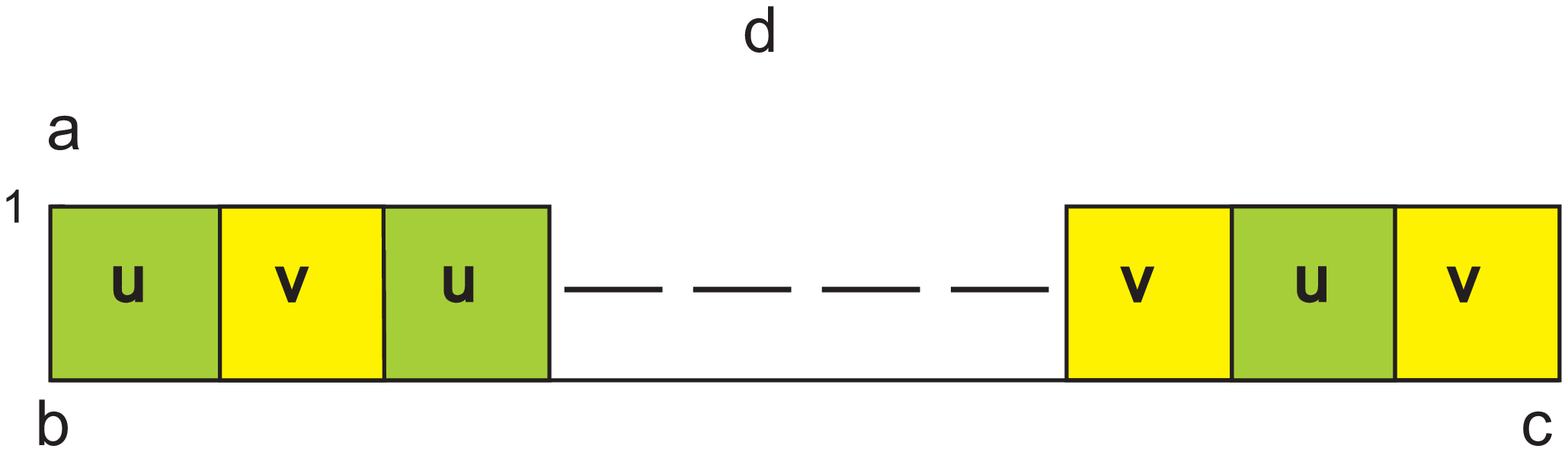}\\
    \label{fig:nblock}
}
\hfill
\subfloat[Blend][$n$-block structure on $\R$ or $\vz{T}_1$]
{
    \psfrag{a}{\Huge{\boldmath $\overbrace{\hspace{5.15 cm}}$\unboldmath}}
    \psfrag{b}{$0$}
    \psfrag{c}{$1$}
    \psfrag{d}{\Large{\boldmath $2n$ \unboldmath}}
    \psfrag{1}{$1$}
    \includegraphics[width=70mm]{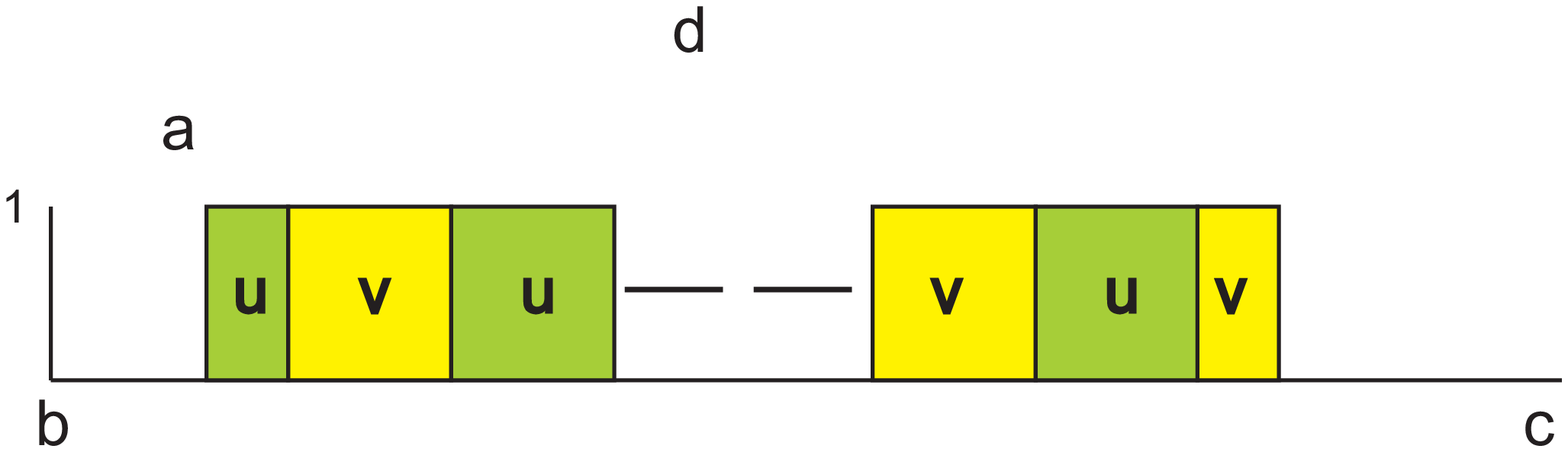}\\
    \label{fig:nblock_blend}
}
\caption{}
\end{figure}

These competing tendencies cause the functional $F_1$ to prefer structures with a specific length scale, as we now illustrate with a simple example in one
space dimension.  For simplicity we take as spatial domain the unit torus $\vz{T}_1$, i.e. the set $[0,1]$ with periodic boundary conditions; all global
minimisers under the condition $u+v\equiv 1$ then are of the form shown in Figure~\ref{fig:nblock}.

For such structures the value of the functional is
\[
F_1 = 2n(c_u+c_v) + \frac1{96n^2},
\]
as can be seen from the results in Section~\ref{sec:globmin1d}.
If we consider $c_u$ and $c_v$ to be fixed, the energy $F_1$ is clearly minimised at a finite value
of $n$. When we study the one-dimensional case on $\R$ without the restriction $u+v\equiv 1$ in more detail, in Section~\ref{sec:globmin1d}, we shall see that the energy actually favours a specific \emph{block width} rather than a specific number of blocks.

\subsection{Blends of co- and homopolymers}

For $u+v\not\equiv1$,  $F_1$ is a model for \emph{blends}, mixtures of
diblock copolymers and homopolymers; the homopolymer is considered to fill the
space not occupied by the diblock copolymer and has local volume fraction $1-u-v$.

The inclusion of homopolymers into a block copolymer melt opens the possibility
of structures with two distinct length scales. The repulsion between the two
blocks creates micro-phase separation at the length scale of the polymer, as described above. At a
larger length scale structures are observed in which regions of pure homopolymer
and pure copolymer alternate.

Blend systems show a tremendous wealth of behaviour. For instance, many different types of
macrodomain geometry have been observed: spheres~\cite{KoizumiHasegawaHashimoto94,OhtaNonomura97,UneyamaDoi04,ZhangJinMa05}, cylinders~\cite{KinningWineyThomas88}, dumbbells~\cite{OhtaIto95}, helices~\cite{HashimotoMitsumuraYamaguchiTakenakaMoritaKawakatsuDoi01}, labyrinths and
sponges \cite{LoewenhauptSteurerHellmannGallot94,Ito98,OhtaIto95}, ball-of-thread~\cite{LoewenhauptSteurerHellmannGallot94}, and many more. In addition, the
microdomains have varying orientation with respect to this macrodomain geometry.
In many cases the micro- and macrodomain geometry appear to be coupled in ways
that are not yet understood.

There is extensive literature on such blend systems, which is mostly
experimental or numerical. For the
numerical experiments it is \emph{de rigeur} to apply a self-consistent mean
field theory and obtain a generalisation of the Ohta-Kawasaki~\cite{OhtaKawasaki86} model (see e.g.~\cite{NoolandiHong83,OhtaNonomura97,ChoksiRen05}).
Of the resulting model the energy $F_1$ is a sharp-interface limit~\cite{Baldo90,ChoksiRen05}.

At the level of mathematical analysis, however, little is known. What form do
global and local minimisers of $F_1$ take? (Do they even exist? The issue of
existence of global minimisers of $F_1$ on $\R$ is first addressed in this paper.) Does the functional
indeed have a preference for layered structures, as the numerical experiments
suggest? What structure and form can macrodomains have? Can we observe in this
simplified functional $F_1$ the breadth of behaviour that is observed in
experiments? All these questions are open, and in this paper we provide some
first answers.

\subsection{Results: global minimisers in one dimension under constrained mass}

The first part of the paper focuses on the one-dimensional situation.

\subsubsection{Existence}
The existence of global minimisers under the constraint of fixed mass follows
mostly from classical arguments (proof of Theorem~\ref{th:exist_real_line}). The non-compactness of the set $\R$ can be
remedied with the cut-and-paste techniques that we introduce to study non-uniqueness (see below).

One non-trivial issue arises when e.g. $c_0=c_u=0$, in which case the functional $F_1$ provides no control on the regularity of $u$. We obtain weak convergence in $L^2$ for a minimising sequence, and therefore a priori we can only conclude that the value set of the limit functions is $[0,1]$, the convex hull of $\{0,1\}$; as a result the limit $(u,v)$ need not be an element of $K_1$. With a detailed study of the stationarity conditions on $u$ we show that stationary points of $F_1$ only assume the extremal values $0$ and $1$. The existence of a minimiser then follows from standard lower semi-continuity arguments.

\subsubsection{Characterisation of macrodomains}
In the one-dimensional situation a macrodomain is a finite sequence of
alternating U- and V-`blocks' or `layers' as in Figure~\ref{fig:nblock_blend}. Choksi and Ren~\cite{ChoksiRen05} studied such macrodomains defined on the torus $\vz{T}_L$ of length $L
$, but their techniques apply unchanged to the real
line also. They showed that if such a macrodomain is stationary, then all \emph{interior} blocks have equal width, while the end blocks are thinner.

\begin{figure}[htb]
\centering
    \psfrag{1}{$1$}
    \psfrag{2}{$2$}
    \psfrag{3}{$3$}
    \psfrag{4}{$4$}
    \psfrag{5}{$5$}
    \psfrag{6}{$6$}
    \psfrag{7}{$7$}
    \psfrag{8}{$8$}
    \psfrag{a}{$a$}
    \psfrag{b}{$b$}
    \psfrag{c}{$c$}
    \psfrag{d}{$d$}
    \psfrag{e}{$e$}
    \psfrag{f}{$f$}
    \psfrag{x}{$M$}
    \psfrag{y}{$\frac{F_1}{M}$}
    \includegraphics[width=100mm]{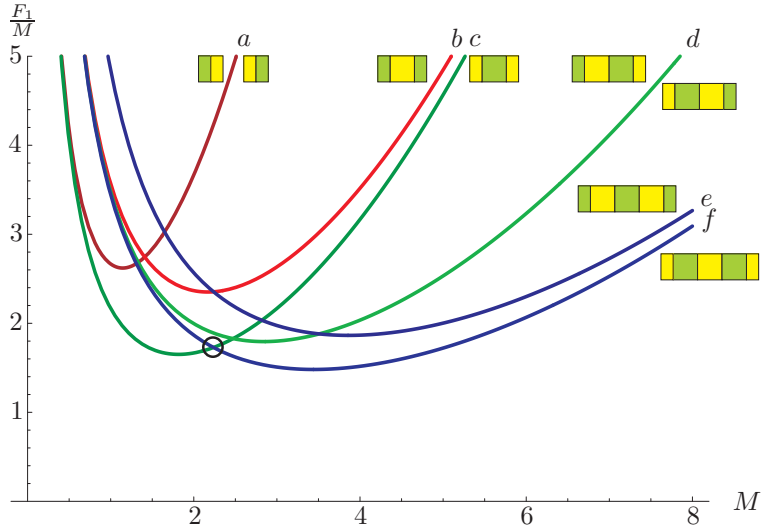}\\
    \caption{Energy per unit mass for the one-dimensional case $\R$,
according to the calculations in Section~\ref{sec:lower_bnd_1d}. $M$ is the
total U-mass; for the surface tension parameters (see Lemma~\ref{lemma:d_ij}) the values $d_{u0} = 1, d_{uv} = 0.7$ and $d_
{v0} = 0.3$ are chosen. The graphs belong to the following structures, as indicated in the figure as well (the lighter coloured blocks are V-blocks, the darker ones U-blocks): (a) UV and VU, (b) UVU, (c) VUV, (d) UVUV and VUVU, (e) UVUVU, (f) VUVUV. The circle indicates where the optimal structure changes.}\label{fig:blocks_2x3cases-intro}
\end{figure}

The exact dimensions of the blocks are fully determined by the number of blocks,
the total mass, and, in the case of the torus, the size of the domain (see Theorems~\ref{th:CR} and~\ref{th:CR2}). It is instructive to minimise $F_1$ within classes defined by a
specific choice of the sequence of U- and V-blocks; Figure~\ref{fig:blocks_2x3cases-intro} shows this minimal energy for different classes and different values of the mass.

\subsubsection{Characterisation of constrained minimisers}
We extend the results of Choksi \& Ren into a full characterisation of global
minimisers, by showing that there exists a global minimiser with only one
macrodomain, and by fully characterising all \emph{other} global minimisers
in terms of the parameters and the morphology (Theorem~\ref{th:exist_real_line}).

This characterisation shows that {non-uniqueness} of minimisers can take two
different forms. The first is the possibility that two different UV-sequences
with the same mass have the same energy, as is illustrated by the encircled
intersection in Figure~\ref{fig:blocks_2x3cases-intro}. This is a common
occurrence in variational problems, where a parameter change causes the global minimum to switch from one local minimiser to another.

The second form of non-uniqueness is related to the fact, which we prove in
Section~\ref{subsec:connected_support}, that two separate macrodomains can be
translated towards each other and joined together without increasing the energy.
In fact, in many cases the energy strictly decreases, and it is this possibility
of strict decrease that allows us to rule out many cases. This leaves us with a set of
conditions for the case of unchanged energy that must be fulfilled when a non-unique global minimiser contains more than one
macrodomain (see Theorem~\ref{th:exist_real_line}). This type of non-uniqueness is specific for the problem
at hand, and produces not a discrete set of minimisers but a continuum,
parametrised by the spacing between the macrodomains.

Although the focus of this paper lies on the unbounded domains $\R$ and $\R^N$, we make a brief excursion to extend the characterisation of global minimisers to the case of the torus $\vz T_L$ with length $L$ (Theorem~\ref{th:exist_periodic_1d}).

\subsubsection{A lower bound}

Figure~\ref{fig:blocks_2x3cases-intro} and more clearly Figure~\ref
{fig:blocks_asymp} illustrate that as the imposed mass increases the number of
blocks of the global minimiser(s) also increases. In Section~\ref{sec:lower_bnd_1d} we calculate values of the energy for various global minimisers, and show that the
thickness of the internal layers approaches the optimal spacing of
\[
2m_0 := 6^{1/3}(c_u+c_v)^{1/3},
\]
for $M \to \infty$ while the width of the end layers converges to half this value (Remark~\ref{rem:convergenceofwidth}).

As a corollary we obtain an {explicit} and {sharp} lower bound for
the energy on $\R$ (Theorem~\ref{th:lower_bound_1d}):
\begin{equation}
\label{est:lower_bound_1d_intro_0}
F_1 (u,v) \geq 2(c_0+\min(c_u,c_v))
  + \left(\frac92\right)^{1/3}(c_u+c_v)^{2/3}\int_\R u.
\end{equation}
The fact that the lower bound is sharp is significant. For instance, the affine
dependence of the right-hand side on the mass $\int u$ implies that the minimal
energy per unit of mass, $F_1(u,v) \left(\int u\right)^{-1}$, is generically not
attained at any finite mass, but only in the limit $\int u \to \infty$.

The word `generic' refers here to the assumption that $c_0+\min(c_u,c_v)>0$, and
the alternative case $c_0=c_u=0$ (or $c_0=c_v=0$) is fundamentally different. In
this latter case macrodomains can be split and joined without changing the
energy.

The characterisation of global minimisers also allows us to establish an asymptotically sharp upper bound (Theorem~\ref{th:lower_bound_1d}):
\begin{equation}
\label{est:upper_bound_1d}
\lim_{M\to\infty} \inf\left\{M^{-1} F_1(u,v): (u,v)\in K_1,
    \ \int u = M\right\} = \left(\frac92\right)^{1/3}(c_u+c_v)^{2/3}.
\end{equation}
In the limit $M\to\infty$, the bound~\pref{est:lower_bound_1d_intro_0} coincides with~\pref{est:upper_bound_1d}.

\subsection{Results: higher dimensions}

\subsubsection{Energy bounds}
A common strategy in the study of pattern-forming systems is not to make any
\emph{Ans\"atze} about the morphology but to search for \emph{weaker
characterisations of behaviour}. As an example of this in the field of block
copolymers, Choksi proves that for pure diblock melts the energy is bounded from
below  by a lower bound with a certain scaling in terms of the physical
parameters---without making any \emph{a priori} assumptions on the morphology~\cite{Choksi01}. This scaling is shared by periodic lamellar structures with a
specific lamellar separation.

For the case at hand, the one-dimensional analysis provides both a lower and an
upper bound on the energy in one dimension. Weakening the lower bound~\pref
{est:lower_bound_1d_intro_0} to
\begin{equation}
\label{eq:lower_bound_intro}
F_1(u,v) \geq \left(\frac92\right)^{1/3}(c_u+c_v)^{2/3} \int u,
\end{equation}
one might conjecture that the lower bound~\pref{eq:lower_bound_intro} holds in $
\R^N$, again without making any \emph{a priori} assumption on the morphology.
However, we have no proof of this conjecture, and in fact, results on mono- and
bilayer stability (see Section~\ref{subsubsec:companion} below) suggest that
such a conjecture may only hold for certain choices of the parameters. In
Section~\ref{sec:scaling} we instead prove a lower bound which is also linear in
mass, but has a smaller constant (Theorem~\ref{lem:umassineqs}).

The explicit construction used to prove the upper bound~\pref{est:upper_bound_1d} suggests a natural strategy for proving a similar upper bound in $\R^N$. In Sections~\ref{sec:upper_bound} and~\ref{subsec:examples} we extend one-dimensional minimisers as lamellar structures in $\R^N$, and prove the same upper bound~\pref{est:upper_bound_1d} in $\R^N$. Here the main step in
the proof is the estimation that `boundary effects' as a result of the cutoff to
finite mass are of lower order.

In Section~\ref{subsec:examples} we use the same idea to calculate the energy values of some
structures with spherical geometry: either solid spheres of one phase (U or V) surrounded
by a spherical layer of the other phase  \emph{(micelles)}, or ring-shaped layered
structures. In both cases the asymptotic energy  exceeds that given by the upper bound~\pref{est:upper_bound_1d}, indicating that they can not be global minimisers in~$\R^N$.

\subsubsection{Monolayer and bilayer stability in periodic strips}
\label{subsubsec:companion}
In a companion paper~\cite{vanGennipPeletier07b} we study the stability with respect to a certain class of perturbations of monolayers and bilayers,
i.e. straight layered structures with one respectively two lines of U-V
interface, in a periodic strip $\vz{T}_L\times\R$. There we show that for
sufficiently large $L$ a monolayer (the simplest lamellar structure, of the form
UV) is always unstable, while the stability of a bilayer (UVU or VUV) depends on the
parameters. For the case of a UVU bilayer with optimal thickness, for instance,
we prove a stability criterion of the form
\[
\text{stability}\qquad \Longleftrightarrow \qquad \frac{c_u+c_v}{c_0+2c_u+c_v} \geq
g\left(\frac{L}{(c_0+2c_u+c_v)^{1/3}}\right),
\]
where $g$ is a continuous function with values in $(0,1)$. Therefore, the bilayer can be stable or unstable,
depending on the relative values in the interface penalisation parameters. Note that the relative value of $c_u+c_v$ should not be too small in order to have stability. More about the special role of $c_u+c_v$ follows in Section~\ref{subsec:d_12}.

\subsection{Related work: partial localisation}
\label{subsec:partloc}

In previous work, one of the authors (Peletier) and R{\"o}ger studied a related
functional whose derivation was inspired by lipid bilayers~\cite{PeletierRoeger06}.
Lipid bilayers might be considered block copolymers, and therefore it is not
surprising that the functional considered in~\cite{PeletierRoeger06} is similar to $F_1
$:
\begin{equation}\label{eq:lipidbil}
\mathcal{F}_{\epsilon}(u, v) := \left\{ \begin{array}{ll}
\displaystyle \epsilon \int_{\R^2} |\nabla u| + \frac{1}{\epsilon} d_1(u, v) & \mbox
{ if $(u, v) \in \mathcal{K}_{\epsilon}$,}\vspace{0.25cm}\\ \infty &\mbox{ otherwise.} \end
{array} \right.
\end{equation}
Here $u$ is the volume fraction or density of lipid heads, $v$ is the volume fraction of
lipid tails, $d_1(\cdot,\cdot)$ is the Monge-Kantorovich distance and
\[
\mathcal{K}_{\epsilon} := \left\{ (u, v) \in \text{BV}(\vz{R}^2; \{0, 1/\epsilon\})^2 : uv = 0 \text{ a.e., and }\int_{\R^2} u = \int_{\R^2} v = M \right\}.
\]
Apart from the choices $c_0 = c_v = 0$ and $c_u = 1$, the main difference
between (\ref{eq:functional}) and (\ref{eq:lipidbil}) is the different non-local
term.

Note that the scaling (constant mass but increasing amplitude $1/\e$) implies
that the supports of $u$ and $v$ shrink to zero measure. The main goal in~\cite{PeletierRoeger06} was to investigate the limit $\e\to0$, and connect the limit
behaviour to macroscopic mechanical properties of the lipid bilayers such as
stretching, bending, and fracture.

The authors studied sequences $(u_\e,v_\e)$ for which the rescaled energy $
\mathcal{G}_{\epsilon} := {\epsilon^{-2}} (\mathcal{F}_{\epsilon} - 2 M)$
remains finite. They revealed a remarkable property of the functional $\mathcal
{G}_{\epsilon}$ (or $\mathcal{F}_{\epsilon}$): boundedness of $\mathcal G_\e(u_
\epsilon,v_\epsilon)$ implies that the support of $u_\e$ and $v_\e$ becomes
close, in the sense of Hausdorff distance between sets, to a collection of
closed curves of total length $M$. The curve-like behaviour indicates \emph
{partial localisation:} localisation in one direction (normal to the limit
curve) and non-localisation in the direction of the tangent. In addition one can
recognise resistance to stretching (because of the fixed length) and resistance
to fracture (because the curves are closed). Moreover, the curves' support is approximately of 'thickness' $2\epsilon$, indicating an underlying bilayer structure. The authors also showed that $\mathcal{G}_{\epsilon}$ Gamma-converges to the {\it Elastica functional}, which
penalises the curvature of curves, showing a tendency of the limit curves to
resist bending.

\medskip

These results suggest considering similar limits for the functional~$F_1$. In
fact the subscript $1$ in $F_1$ and $K_1$ already refers to the appropriate
rescaling:
\begin{equation}
\label{def:Fe}
F_\epsilon(u, v) = \left\{ \begin{array}{ll} \displaystyle
 \epsilon\left(c_0 \int_{\R^N} |\nabla (u + v)| + c_u \int_{\R^N} |\nabla u|
+ c_v \int_{\R^N} |\nabla v|\right)
  +  \frac1\epsilon\|u - v\|_{H^{-1}}^2
  & \mbox{ if $(u, v) \in K_\epsilon$,}\vspace{0.25cm}\\
\infty &\mbox{ otherwise,} \end{array} \right.
\end{equation}
where
\[
K_\epsilon := \left\{ (u, v) \in \left(\text{BV}(\R^N)\right)^2 :
   u(x), v(x) \in \{0, 1/\epsilon\} \text{ a.e., and } uv = 0 \text{ a.e., and }
   \int_{\R^N} u = \int_{\R^N} v \; \right\}.
\]
As mentioned above, in the companion paper~\cite{vanGennipPeletier07b} we investigate the stability
of bilayers, and show that parameter choices exist for which they are stable:
this provides another suggestion that the functional $F_\epsilon$ may display
similar behaviour in the limit $\epsilon\downarrow0$. This is work for future research.

\section{Preliminary definitions}

\subsection{Problem setting}
In this paper we mostly consider as domain the whole space $\R^N$; however, sometimes we will make an excursion to the
torus $\vz{T}_L^N$, i.e. a periodic cell $\prod_{i=1}^N [0,L_i]$ with the endpoints of each interval identified.

\begin{definition}
For $f\in L^1(\R^N)$ (or $L^1(\vz T_L^N)$) with $\int f =0$ and compact support,
\begin{equation}
\label{def:HMO}
\|f\|_{H^{-1}}^2 := \int fG*f,
\end{equation}
where $G$ is a Green's function of the operator $-\Delta$ on $\R^N$ (or on $\vz T_L^N$). We define the space $H^{-1}(\R^N)$ as the completion of $\left\{f\in L^1(\R^N): \supp f \text{ compact}, \int_{\R^N} f=0\right\}$ with respect to the norm in (\ref{def:HMO}). Similarly $H^{-1}(\vz{T}_L^N)$ is defined as the completion of $\left\{f\in L^1(\vz{T}_L^N): \int_{\vz{T}_L^N} f=0\right\}$ with respect to this norm.
\end{definition}

On $\vz{T}_L^N$ the zero average condition of $f$ is necessary in order for $G*f$ to respect the topology of the torus:
\[
\int_{\vz{T}_L^N} f = -\int_{\vz{T}_L^N} \Delta(G*f) = 0.
\]
This condition also allows for a convenient reformulation of the norm~\pref{def:HMO} in terms of the \emph{Poisson potential} $\phi_f$ of $f$, given by
\[
\phi_f = G*f,
\]
such that
\begin{equation}
\label{eq:calcHMO}
\|f\|_{H^{-1}}^2 = \int f\phi_f = \int |\nabla\phi_f|^2.
\end{equation}
In some cases it will be useful to add a constant to $\phi_f$; note that this can be done without changing the value in~\pref{eq:calcHMO}.

If the set $H^1_0$ is defined as the completion of $C_c^1(\R^N)$ (or $C^1(\vz T_L^N)$ with zero mean) with respect to the norm $\|g\|_{H^1_0}^2=\int|\nabla g|^2$, then~\pref{def:HMO} is the dual norm of $H^1_0$ with respect to the $L^2$-inner product and satisfies
\[
\int fg \leq \|f\|_{H^{-1}} \|g\|_{H^1_0},
\]
for all $f\in H^{-1}$ and $g\in H_0^1$.

We repeat the definition of $F_1$ and $K_1$ for convenience.

\begin{definition}\label{def:functional}
Let $c_0$, $c_u$, and $c_v$ be real numbers.
Define
\[
\label{eq:functional2}
F_1(u, v) = \left\{ \begin{array}{ll}
\displaystyle c_0 \int_{\R^N} |\nabla (u + v)| + c_u \int_{\R^N} |\nabla u|
+ c_v \int_{\R^N} |\nabla v|\hspace{0.25cm}
  +  \|u - v\|_{H^{-1}}^2
& \mbox{ if $(u, v) \in K_1$,}\vspace{0.25cm}\\
\infty &\mbox{ otherwise,} \end{array} \right.
\]
where the admissible set is given by
\[
K_1 := \left\{ (u, v) \in \left(\text{BV}(\R^N)\right)^2 :
   u(x), v(x) \in \{0, 1\} \text{ a.e., and}\ uv = 0 \text{ a.e., and } \int_{\R^N}
u = \int_{\R^N} v\right\}.
\]
\end{definition}
We will require $c_0, c_u$ and $c_v$ to be non-negative and assume that at least one of these coefficients is positive. See also Remark~\ref{rem:coefficients}.

Sometimes we consider the case of the torus instead of $\R^N$. It is understood that in the above definition the instances of $\R^N$ are then replaced by $\vz{T}_L^N$.

Another, equivalent, form of the functional will be useful, in which the
penalisation of the three types of interface U-0, V-0, and U-V, is given explicitly by surface tension
coefficients $d_{kl}$:
\begin{lemma}
\label{lemma:d_ij}
Let the \emph{surface tension coefficients} be given by
\begin{align*}
d_{u0} &:= c_u+c_0,\\
d_{v0} &:= c_v+c_0,\\
d_{uv} &:= c_u+c_v,
\end{align*}
Non-negativity of the $c_i$ is equivalent to the conditions
\footnote{The indices $j, k, l$ take values in $\{u, v, 0\}$ and the $d_{kl}$
are taken symmetric in their indices, i.e. $d_{vu} := d_{uv}$ etc.}
\begin{equation}\label{eq:ddemands}
0 \leq d_{kl} \leq d_{kj} + d_{jl} \qquad\text{for each } k\not=l.
\end{equation}

Then
\[
F_1(u, v) = \left\{ \begin{array}{ll}
  d_{u0}\HNo(S_{u0}) + d_{v0}\HNo(S_{v0}) + d_{uv}\HNo(S_{uv})
+  \|u - v\|_{H^{-1}}^2
& \mbox{ if $(u, v) \in K_1$,}\\ \infty &\mbox{ otherwise.} \end{array} \right.
\]
where $S_{kl}$ is the interface between the phases $k$ and $l$:
\begin{align*}
&S_{u0} = \partial^* \supp u \setminus \partial^* \supp v,\\
&S_{v0} = \partial^* \supp v \setminus \partial^*\supp u,\\
&S_{uv} = \partial^* \supp u \cap \partial^* \supp v,
\end{align*}
and $\partial^*$ is the essential boundary of a set.
\end{lemma}

\begin{remark}
The essential boundary of a set consists of all points in the set that have a density other than $0
$ or $1$ in the set. Details can be found in \cite[Chapter 3.5]{AmbrosioFuscoPallara00}.
\end{remark}

\begin{proof}[Proof of Lemma~\ref{lemma:d_ij}]
The main step in recognising the equivalence of both forms of $F_1$ is noticing
that, if $u$ is a characteristic function, then
\[
\int |\nabla u| = \HNo(\partial^* \supp u \cap \Omega).
\]
\end{proof}
Note the different interpretations of the coefficients $c_i$ and the surface tension coefficients $d_{kl}$. The latter have a direct physical interpretation: they determine the mutual repulsion between the different constituents of the diblock copolymer-homopolymer blend. For example, the value of $d_{uv}$ (as compared to the values of $d_{u0}, d_{v0}$ and $1$, the coefficient in front of the $H^{-1}$-norm) determines the energy penalty associated with close proximity of U- and V-polymers. In particular, if one of these surface tension coefficients is zero, the corresponding polymers do not repel each other and many interfaces between their respective phases in the model can be expected. On the other hand the coefficients $c_i$, when taken separately, do not convey complete information about the penalisation of the boundary of a phase. If for instance $c_u=0$, but $c_v\neq0$, the part of the U-phase interface that borders on the V-phase still receives a penalty, because $d_{uv}=c_v$. For this reason the use of surface tension coefficients makes more sense from a physical point of view. For the mathematics it is often easier to use the formulation in terms of $c_i$.

\begin{remark}\label{rem:coefficients}
The condition~\pref{eq:ddemands} can be understood in several ways. If, for
instance, $d_{uv}>d_{u0}+d_{v0}$, then the U-V type interface, which is penalised
with a weight of $d_{uv}$, is unstable, for the energy can be reduced by
slightly separating  the U and V regions and creating a thin zone of 0
inbetween. A different way of seeing the necessity of~\pref{eq:ddemands} is by
remarking that the equivalent requirement of non-negativity of the $c_i$ is
necessary for $F_1$ to be lower semicontinuous in e.g. the $L^1$ topology. Our assumption that at least one $c_i$ is positive is equivalent to assuming that at least two $d_{kl}$ are positive.
\end{remark}

\subsection{The role of $d_{uv}$}
\label{subsec:d_12}

The behaviour of the model described by $F_1$ is crucially different in the two
cases $d_{uv}>0$ ($c_u+c_v>0$) and $d_{uv}=0$ ($c_u=c_v=0$). The statements made in the introduction such as
'the functional $F_1$ prefers structures with a definite length scale' actually
only hold in the case $d_{uv}>0$.
For most results in this work we will assume this condition to hold, and to justify this we now show with an example how the case $d_{uv}=0$ is
different.

Consider the one-dimensional case, take $\Omega$ to be the torus $\vz{T}_1$, and
fix $c_0=1$ and $c_u=c_v=0$, or equivalently $d_{uv}=0$ and $d_{u0}=d_{v0}=1$.
Restricting ourselves to functions $(u,v)\in K_1$ with $\int_0^1 u = \int_0^1 v = M$,
for some fixed mass $0<M<1/2$, we find that for any $(u,v)$ there are at least
two U-0 or V-0 type transitions, and therefore
\[
F_1(u,v) = \int_0^1 |(u+v)'| \hspace{0.2cm}+ \|u-v\|_{H^{-1}}^2
  \geq 2.
\]
On the other hand, equality is only reached if $u-v=0$, which is not possible
for positive mass~$M$. But the value $2$ can be reached by a sequence of
approximating pairs $(u_n,v_n)$,
\begin{align*}
&u_n(x) = \begin{cases}
1 & |x|\leq n \text{ and }
  \frac {2k}{n} < x < \frac{2k+1}n, \text{ for some } k\in \mathbb Z \\
0  &\text{otherwise}\end{cases}\\
&v_n(x) = \begin{cases}
1 & |x|\leq n \text{ and }
  \frac {2k-1}{n} < x < \frac{2k}n, \text{ for some } k\in \mathbb Z \\
0  &\text{otherwise}\end{cases}
\end{align*}
Then $(u_n,v_n)\in K_1$ and
\begin{itemize}
\item $\int_0^1 |(u_n+v_n)'| = \int_0^1 |\chi_{[-n,n]}'| = 2$;
\item In Section~\ref{sec:globmin1d}  it is calculated that a single one-dimensional monolayer of width $2m$ and height $1$ satisfies $\|u - v\|_{H^{-1}}
^2 = 2m^3/3$; extending this result to the functions $(u_n,v_n)$, which are

concatenations   of $n^2$ such monolayers, each of width $2/n$, we find $\|u - v\|_
{H^{-1}}^2 = n^2 \cdot 2n^{-3}/3 =2n^{-1}/3$.
\end{itemize}
Consequently, $F_1(u_n,v_n)$ converges to $2$ for $n\to \infty$.

\drop{
\begin{align*}
u_n(x) &= \begin{cases}
1 & \text{if $\ds x\in \left[\frac{4mk}{2n},\frac{2m(2k+1)}{2n}\right)$ for some
$0\leq k\leq n-1$}\\
0 & \text{otherwise},
\end{cases} \\
v_n(x) &= \begin{cases}
1 & \text{if $\ds x\in \left[\frac{2m(2k+1)}{2n},\frac{2m(2k+2)}{2n}\right)$ for
some $0\leq k\leq n-1$}\\
0 & \text{otherwise}.
\end{cases}
\end{align*}
}

This sequence illustrates the preferred behaviour when $d_{uv}=0$: since the
interfaces between the U- and V-phases are not penalised, rapid alternation of U-
and V-phase effectively eliminates the $H^{-1}$-norm, reducing the energy to
the interfacial energy associated with a single field $u+v$.

\section{Global minimisers in one dimension}
\label{sec:globmin1d}
In this section we fully characterise the set of global minimisers of $F_1$ in
one space dimension, i.e. $N=1$. Our main discussion concerns the case of $\R$,
but in Section~\ref{subsec:ex_on_TL} we will briefly mention results on the torus $\vz{T}_L$.

In one space dimension it is useful to regard admissible functions $(u,v)$ as
a sequence of \emph{blocks}. A \emph{U-block}, a \emph{V-block}, and a \emph{0-block} are connected components of $\supp u$, $\supp v$,  and $\R\setminus
\supp(u+v)$, respectively. Adjacent blocks are separated by \emph{transitions}
or \emph{interfaces}. We will see below (Corollary~\ref{corol:finite-interfaces}) that any stationary point has a finite number of interfaces, even
if either $d_{u0}$ or $d_{v0}$ vanishes.

If $(u, v)$ is an admissible pair, each of the connected components of its
support $\supp(u+v)$ is in fact a macrodomain in the sense of the introduction.
If there is only one such macrodomain, we call the configuration \emph{connected}. Thinking about the structures in terms of sequences of blocks, we can specify connected configurations up to block width and translation by a sequence of U's and V's, e.g. UVUVU.

Characterising the set of global minimisers falls apart into two steps:
\begin{itemize}
\item[A] For a given macrodomain we describe the optimal spacing between the
transitions;
\item[B] We derive necessary conditions for the occurrence of a disconnected global minimiser, i.e. a global minimiser with more than one macrodomain.
\end{itemize}
In addition we use the techniques of part B above to prove the existence of a global minimiser.

In Section~\ref{subsec:macrodomains} we first describe the characterisation given
by Choksi and Ren~\cite{ChoksiRen05} of the internal structure of macrodomains, which
essentially coincides with part A above. We then continue in Section~\ref
{subsec:connected_support} by showing that the support can be reduced to a
single connected component; this also provides necessary and sufficient
conditions for non-uniqueness (Theorem~\ref{th:exist_real_line}).
The reduction to a single macrodomain also allows us to prove an existence
result (Theorem~\ref{th:exist_real_line}).
Finally, in Section~\ref{sec:lower_bnd_1d}, we calculate the values of these
minimisers and derive a lower bound for the energy per unit of mass.

\subsection{Stationarity}
\label{subsec:Stationarity}

Because the set of admissible functions $K_1$ is not locally convex we need to carefully formulate the notion of stationary point.

\begin{definition}
We call $(u, v) \in K_1$ a stationary point of $F_1$ if for any sequence $(u_n, v_n)\subset K_1$ such that  $u_n\to u$ in $L^1$ and $v_n\to v$ in $L^1$,
\[
|F_1(u, v) - F(u_n, v_n)| = o\left(\int_{\Omega} |u-u_n|\,dx + \int_{\Omega} |v-v_n|\,dx\right).
\]
\end{definition}
As a consequence of this definition, if $t\mapsto (u(t), v(t))$ is a curve in $K_1$, with $(u(0), v(0))$ a stationary point of $F_1$, then
\[
\left.\frac{d}{dt} F_1(u(t), v(t))\right|_{t=0} = 0.
\]
In the proofs of the results in Section~\ref{subsec:macrodomains} a special case of this is used: for a connected configuration in one dimension that is stationary under constrained mass the derivative of $F_1$ with respect to mass-preserving changes in the position of the interfaces is zero.

\subsection{Characterisation of macrodomains}
\label{subsec:macrodomains}

For periodic domains, Choksi and Ren~\cite{ChoksiRen05} have given a characterisation of the structure of macrodomains. For its formulation it is useful to define three \emph{types} of interface. Interfaces 0-U and U-0 interfaces are considered to be of the same type, as are 0-V and V-0 interfaces and U-V and V-U interfaces. Choksi and Ren's conclusions are
\begin{theorem}[\cite{ChoksiRen05}]
\label{th:CR}
Let $(u,v)$ be a stationary point of $F_1$ on the torus $\vz T_L$ under constrained mass, with
$\supp(u+v)$ connected and with a finite number of interfaces. Then
\begin{enumerate}
\item\label{thCR:1} Each pair of adjacent U-V type transitions is separated by the
same amount; i.e. each U- or V-block is of the same width, with the exception of
the two end blocks.
\item\label{thCR:2} In the cases UVUV\ldots U and VUVU\ldots V the end blocks
are half as wide as the internal blocks.
\item\label{thCR:3} In the case UVUV\ldots V (or the mirrored configuration VUVU\ldots U) there is an additional relation
that determines the width of the end blocks.
\end{enumerate}
\end{theorem}

The case of $\R$ was not explicitly discussed by Choksi and Ren, but both the
result and the proof for this case are simpler than for the periodic cell:
\begin{theorem}
\label{th:CR2}
Let $(u,v)$ be a stationary point of $F_1$ on $\R$ under constrained mass, with $\supp(u+v)$ connected and with a finite number of
interfaces. Then
\begin{enumerate}
\item\label{thCR2:1} Each pair of adjacent U-V type transitions is separated by the
same amount; i.e. each U- or V-block is of the same width, with the exception of
the two end blocks.
\item\label{thCR2:2} The end blocks are half as wide as the internal blocks.
\end{enumerate}
\end{theorem}

The main tool in the proof of these theorems is the following lemma.

\begin{lemma}[{\cite[Lemma 4.1]{ChoksiRen05}}]
\label{lemma:equal_phi}
For any stationary point under constrained mass,
the Poisson potential $\phi$ has equal value at any two interfaces of the same {type}.\end{lemma}
The statements about the block sizes are deduced from this lemma, and from the
fact that the potential $\phi$ has prescribed second derivative on each block.

\subsection{Reduction to connected support}
\label{subsec:connected_support}

We first need a technical result to rule out the possibility of an infinity of
transitions.

\begin{lemma}
Let $(u,v)$ be a stationary point under constrained mass, let $\Omega$ be either
$\R$ or $\vz{T}_L$ and let $\omega\subset \Omega$ be an open set such that $v
(\omega) = \{0\}$. Then $\omega$ contains at most two U-0 type transitions. A
similar statement holds with $u$ and $v$ exchanged.
\end{lemma}

\begin{proof}
On $\omega$, $\phi''\leq0$; each U-0 or 0-U transition occurs at the same value
of $\phi$ (Lemma~\ref{lemma:equal_phi}), say at $\phi=c\in\R$. If the set $\{x
\in\omega:\phi(x) = c\}$ has more than two elements, then by convexity,
\begin{align*}
&\phi(x)= c \qquad\text{for }x\in [x_1,x_2],\\
&\phi(x) < c \qquad\text{for }x\in \omega\setminus[x_1,x_2],
\end{align*}
for some $x_1 < x_2\in\omega$.
On $(x_1,x_2)$, therefore, $\phi''=0$ and thus $u=0$. Therefore there are at
most two transitions connecting U and 0, at $x=x_1$ and at $x=x_2$.
\end{proof}

\begin{corol}
\label{corol:finite-interfaces}
If $d_{uv}>0$, then a stationary point under constrained mass has a finite
number of transitions.
\end{corol}

\begin{proof}
By~\pref{eq:ddemands}, at least two out of the three $d_{ij}$ are strictly
positive. If all three are positive, then the finiteness of $F_1$ implies a
bound on the number of interfaces. If one is zero, say $d_{u0}$, then the lemma
above states that the number of U-0 or 0-U transitions is no larger than the
number of V-interfaces. Since the latter is bounded, the former is also.
\end{proof}

\begin{theorem}
\label{th:exist_real_line}
Let $N=1$. Let $d_{uv} > 0$, and fix a mass $M>0$.
\begin{enumerate}
\item There exists a global minimiser under constrained mass $M$ for which $
\supp(u+v)$ is connected.\label{item:reallineconnmin}
\item This global minimiser is \emph{non}-unique (apart from translation and
mirroring) if and only if
  \begin{enumerate}
    \item the energy of this configuration is equal to the energy of another
configuration $(\bar u, \bar v)$ for which $\supp(\bar u+ \bar v)$ is also
connected, or \label{item:sameenergy2}
    \item one of the following two conditions is satisfied:
      \label{item:sameenergy3}
        \begin{enumerate}
      \item $d_{u0}=0$ and there exists a global minimiser with an internal U-block or;\label{item:RL:condunique1}
      \item $d_{v0}=0$ and there exists a global minimiser with an internal V-block.\label{item:RL:condunique2}
    \end{enumerate}
  \end{enumerate}
\end{enumerate}
\end{theorem}

The non-uniqueness mentioned in condition~\ref{item:sameenergy} can manifest
itself in multiple ways. Figure~\ref{fig:blocks_2x3cases-intro} shows how the
optimal structure varies with mass: as the mass increases, the global minimiser
progresses through structures with more and more layers. At the intersection
points of the curves in the figure, indicated by a circle, structures belonging
to different curves have the same value of the energy. Another possibility
occurs when $d_{u0}=d_{v0}$, since then $u$ and $v$ can be interchanged without
changing the energy. The situation where two minimisers are both connected, have
the same sequence of blocks (up to mirroring), but differ in the block widths,
however, is ruled out by Theorem~\ref{th:CR2}.

The fact that the global minimiser can be non-unique when, for example, $d_{u0}
=0$ can easily be recognised by an example. Suppose that there exists a global
minimiser of the form UVUVU. Since the outer blocks of this structure are both
U-blocks, Lemma~\ref{lemma:equal_phi} states that the value of $\phi$ is the
same at the two interfaces of U-0 type, and $\phi$ is therefore symmetric around
the middle of the structure.

We now split the structure at the middle into two parts, and move the two parts
apart. In doing so we create two new U-0 type transitions, which carry no energy
penalty since we assumed $d_{u0}=0$. Since we split at the middle, where $
\phi'=0$, the new potential $\phi$ can be constructed from the old one by
translation of the parts, and the value of $\|u-v\|_{H^{-1}}$ is also unchanged.

\begin{proof}[Proof of Theorem~\ref{th:exist_real_line}]
We defer the proof of existence of a global constrained minimiser to the end,
and start by showing that existence of a global minimiser implies existence of
a global \emph{connected} minimiser.

Suppose $(u, v) \in K_1$ is a global minimiser such that $\R\setminus \supp(u+v)
$ has at least three connected components. By Corollary~\ref{corol:finite-interfaces} the support of $u+v$ is bounded, and therefore we can take those
three components to be $(-\infty,0)$, $(x_1,x_2)$, and $(x_3,\infty)$. The
points $0$, $x_1$, $x_2$, and $x_3$ therefore all are interfaces.

Let $\phi$ be the associated potential; since $u$ and $v$ vanish on $(x_1,x_2)$
and $(x_3,\infty)$, $\phi$ is linear on $(x_1,x_2)$ and constant on $(x_3,
\infty)$. Denote by $\phi'_{12}$ the value of  $\phi'(x)$  for $x \in [x_1, x_2]
$.

For any $0<a\leq {x_2-x_1}$, which we fix for the moment, we construct a new
pair of functions $\bar u$ and $\bar v$ with associated potential $\bar \phi$ as
follows. Set
\begin{align}
\label{def:overline_u}
\bar u(x) &:= \begin{cases}
  u(x) & x\leq x_1\\
  u(x+a) & x_1<x<x_3-a\\
  0 & x\geq x_3-a
\end{cases}\\
\bar v(x) &:= \begin{cases}
  v(x) & x\leq x_1\\
  v(x+a) & x_1<x<x_3-a\\
  0 & x\geq x_3-a
\end{cases}\\
\widetilde\phi(x) &:= \begin{cases}
  \phi(x) & x\leq x_1\\
  \phi(x+a) -\phi(x_1+a) + \phi(x_1) & x_1<x<x_3-a\\
  \phi(x_3)-\phi(x_1+a)+\phi(x_1) & x\geq x_3-a
\end{cases}
\label{def:widetilde_phi}
\end{align}
Because $\phi'(x_1) = \phi'(x_1+a) = \phi'_{12}$, the function $\widetilde \phi$
is continuously differentiable on $\R$; and since $\widetilde\phi$ satisfies $
{\widetilde \phi}'' = \bar u - \bar v$
on $\R$, it is the Poisson potential associated with $\bar u$ and $\bar v$.

We now show that $F_1(\bar u,\bar v)\leq F_1(u,v)$.
As for the interfacial term in $F_1$, if $0<a<x_2-x_1$, then the various
transitions remain the same, only translated to different positions; therefore
the interfacial term is unchanged. In the case $a=x_2-x_1$, in comparison with $
(u,v)$ the two interfaces at $x=x_1$ and $x=x_2$ have been joined to one
interface, or have even annihilated each other; by the assumption~\pref
{eq:ddemands} this does not increase the interfacial term.

For the second term of $F_1$ we calculate
\begin{align}
\notag
\int_\R \bigl({\widetilde\phi}'\bigr)^2
 &= \int_{-\infty}^{x_1}{\phi'}^2 + \int_{x_1+a}^{x_3} {\phi'}^2 \\
     \label{ineq:phi1phi3}
 &\leq \int_{-\infty}^{x_1}{\phi'}^2 + \int_{x_1+a}^{x_3} {\phi'}^2 +
     a{\phi'}_{12}^2 \\
 &= \int_{\R} {\phi'}^2.
 \notag
\end{align}
We conclude that $F_1(\bar u,\bar v)\leq F_1(u,v)$. Since $(u, v)$ is a global
minimiser, we conclude that $F_1(\bar u,\bar v)=F_1(u,v)$ and thus that $(\bar
u, \bar v)$ is another global minimiser. Furthermore by Corollary~\ref
{corol:finite-interfaces}, $\supp(u+v)$ has a finite number of connected
components and thus we can repeat this procedure until only one component
remains. Therefore we have proved that if a global minimiser exists, then there
(also) exists a global minimiser with connected support.

\medskip

Assume now that two global minimisers exist, one of which has connected support.
The other global minimiser, let us call it $(u, v)$, either has connected $\supp
(u+v)$ or disconnected $\supp(u+v)$. In the former case we have proved  part~\ref{item:sameenergy} of the theorem; therefore we now assume the latter case,
and show that this implies part~\ref{item:sameenergy3}.

Since $(u, v)$ has disconnected $\supp(u+v)$, we can apply the construction
above. For a given choice of $a$, we find another configuration $(\bar u, \bar
v)$ with energy equal or less than that of $(u, v)$. Since $(u, v)$ is a global
minimiser, the energy of $(\bar u, \bar v)$ is equal to that of $(u, v)$ and
thus the two inequalities encountered above are saturated. In particular,
\begin{itemize}
\item The joining of the two interfaces surrounding a 0-block does not reduce
the energy;
\item The inequality~\pref{ineq:phi1phi3} is saturated.
\end{itemize}

The saturation of~\pref{ineq:phi1phi3} implies that $\phi'_{12}=0$, and
therefore that $\phi(x_1)=\phi(x_2)$. We now prove that these interfaces are of
the same type, \emph{i.e.} either both U-0 type or both V-0 type transitions.

Suppose not, and to be concrete, suppose that the interface at $x=x_1$ is a V-0
transition, and at $x=x_2$ a 0-U transition. In this paragraph we will
explicitly distinguish between mirrored interfaces of the same type, e.g. U-0
and 0-U. Since $-\phi''=u-v$ and $\phi'(x_1) = \phi'(x_2) = \phi'_{12} = 0$,
there exists a $y_2 > 0$ such that the next transition is at $x_2 + y_2$ and $
\phi$ decreases for $x \in (x_2, x_2 + y_2)$, implying that the next transition
can not be a U-0 transition (which would require the same value for $\phi$ as at
$x=x_2$) but is a U-V transition, with a value of $\phi$ less than $\phi(x_2)$.
The same argument holds for the interface at $x_1$: the previous transition is
at $x_1 - y_1$ for a $y_1 > 0$ and is again a U-V transition, this time with a
value of $\phi$ larger than $\phi(x_1) = \phi(x_2)$. Since two U-V transitions
have a different value of $\phi$, the structure is not stationary, a
contradiction.

Since the interfaces at $x_1$ and $x_2$ are of the same type, a non-changing
interface energy implies that either $d_{u0}=0$ or $d_{v0}=0$, which is the first part of conditions~\ref{item:RL:condunique1} and~\ref{item:RL:condunique2}. Since the construction provides a global minimiser with an internal U-block (if $d_{u0}=0$) or an internal V-block (if $d_{v0}=0$), the second part of these conditions is also satisfied.

\smallskip

We have now proved that existence of a disconnected global minimiser implies
condition~\ref{item:sameenergy3}. The opposite statement, that condition~\ref
{item:sameenergy3} suffices for the existence of a disconnected global
minimiser, follows from splitting any minimiser at a point $x$ inside a U-block
(supposing $d_{u0}=0$) such that $\phi'(x)=0$.

\smallskip
It remains to prove the existence of a global minimiser, and we now turn to this
issue.
Let $(u_n,v_n)$ be a minimising sequence. We first note that the translation
arguments that we used above allow us to reduce an arbitrary minimising sequence
to a minimising sequence whose elements each are connected.
Therefore we may assume that the support of the sequence remains inside some
large bounded set $\Omega\subset \R$, and does not approach the boundary of this set.

Since both $u_n$ and $v_n$ are bounded in $L^\infty(\Omega)$, there exist
subsequences (that we again denote by $u_n$ and $v_n$) such that
\[
u_n \weakstarto u_\infty \qquad\text{and}\qquad v_n\weakstarto v_\infty
\qquad\text{in }L^\infty(\Omega).
\]
Note that this convergence implies that $\int u_\infty = \int v_\infty = M$, since the constant $1$ is an element of~$L^1(\Omega)$. Since $L^2(\Omega)\subset L^1(\Omega)$ we also have
\[
u_n \weakto u_\infty \qquad\text{and}\qquad v_n \weakto v_\infty \qquad\text{in }L^2(\Omega).
\]
The functions $u_\infty, v_\infty$, as the weak-* limits of $u_n, v_n$, take values in the interval $[0,1]$. Thus if we replace $K_1$ in (\ref{eq:functional2}), the definition of $F_1$, by (note the change in value set)
\[
\tilde K_1 := \left\{ (u, v) \in \left(\text{BV}(\Omega)\right)^2 :
   u(x), v(x) \in [0, 1] \text{ a.e., and } uv = 0 \text{ a.e., and } \int_{\Omega}
u = \int_{\Omega} v\right\},
\]
then $(u_\infty, v_\infty)\in \tilde K_1$ and $F_1$ is convex on $L^2(\Omega)$. This implies that the subdifferential of $F_1$ at $(u_\infty, v_\infty)$ is non-empty, i.e. there exist $p_1, p_2\in L^2(\Omega)$ such that
\[
F_1(u_n, v_n) \geq F(u_\infty, v_\infty) + \int_\Omega p_1 (u_n-u_\infty) + \int_\Omega p_2 (v_n-v_\infty).
\]
Weak convergence in $L^2(\Omega)$ now gives us lower semi-continuity with respect to this convergence:
\[
F_1(u_\infty,v_\infty) \leq \liminf_{n\to\infty} F_1(u_n,v_n).
\]
It remains to prove that $u_\infty$ and $v_\infty$ are admissible, {i.e.}
that they take values $0$ and $1$ and that $u_\infty v_\infty = 0$ almost everywhere. In other words, we want to show that not only $(u_\infty, v_\infty)\in\tilde K_1$, but even $(u_\infty, v_\infty)\in K_1$.

By the assumption $d_{uv}>0$ at least one of the coefficients $c_u$ and $c_v$ is
strictly positive. Suppose that $c_u>0$; then the boundedness of $\int |
u_n'|$ implies that the convergence of $u_n$ is strong in $L^1$ and
pointwise almost everywhere~\cite[Theorem 5.2.4]{EvansGariepy92}. Therefore, for any $\psi\in L^\infty(\Omega)$,
\[
\int_\Omega \psi u_\infty v_\infty = \lim_{n\to\infty} \int_\Omega \psi u_n v_n = 0,
\]
implying that $u_\infty v_\infty = 0$. Also the pointwise convergence gives
\[
u_\infty\in \{0,1\} \quad\text{a.e.}
\]
If also $c_v>0$, then the same convergence holds for $v_\infty$, and the proof is
done. If instead $c_0>0$, then the same holds for $u_\infty+v_\infty$, and again the
proof is done. We continue under the assumption that $c_0=c_v=0$.

For the pair $(u_\infty,v_\infty)$ to be admissible, it is necessary that $v_\infty$ takes values in the boundary set $\{0,1\}$ only. This is a consequence of the lemma that we state below.
\end{proof}

\begin{lemma}
Let $c_0=c_v=0$.
If $(u,v)$ minimises $F_1$ among all pairs $(\bar u,\bar v)$ such that
\begin{itemize}
\item $\bar u\in BV(\R;\{0,1\})$ and $\bar v\in BV(\R;[0,1])$;
\item $\bar u\bar v = 0$ a.e. in $\R$;
\item $\int_\R \bar u = \int_\R \bar v = \int_\R u$,
\end{itemize}
then $v(x)\in\{0,1\}$ for almost every $x\in\R$.
\end{lemma}
\begin{proof}
Choose $0<\eta<1/2$ and let $\omega\subset \R$ be the set of intermediate values
\[
\omega = \{ x\in \R: v(x) \in (\eta,1-\eta)\}.
\]
We need to prove that $|\omega|=0$.
Assume that $|\omega|>0$ and define a perturbation
\[
\zeta(x) = (\phi(x)-c)\chi_\omega(x),
\]
where $\phi = \phi_{u-v}$ is the Poisson potential associated with $u-v$, $\chi_
\omega$ is the characteristic function of the set $\omega$, and $c$ is a
constant chosen to ensure that $\int\zeta = 0$. Note that almost everywhere on~$
\omega$ the function $\phi$ is twice differentiable with $\phi''\geq \eta>0$.

Since the pair $(u, v+\e \zeta)$ is admissible for $\e$ in a neighbourhood of
zero,
\[
0 = \left.\frac\partial{\partial\e} F_1(u, v+\e \zeta)\right|_{\e=0}
 = 2\int_\R \zeta \phi = 2\int_\omega (\phi-c)^2,
\]
so that $\phi$ is constant a.e. on $\omega$. As $\phi$ is defined up to addition of
constants we may choose $\phi=0$ on~$\omega$.

Since $|\omega|>0$, we can choose $x_0\in\omega$ such that $\omega$ has density
$1$ at $x_0$ and that $\phi$ is twice differentiable at $x_0$, with $\phi''(x_0)
\in(\eta,1-\eta)$. Because of the density condition it is possible to find
sequences $a_n \in \R$, $n\in\N$,  with the properties
\begin{itemize}
\item $a_n \to 0$ as $n\to\infty$;
\item For each $n\in \N$, $x_0\pm a_n \in\omega$.
\end{itemize}
Then
\[
\phi''(x_0) = \lim_{n\to\infty}
  |a_n|^{-2}\bigl[\phi(x_0-a_n) - 2\phi(x_0) + \phi(x_0+a_n)\bigr]
  = 0,
\]
a contradiction with $\phi''(x_0)\geq \eta$, and therefore with the assumption
that $\omega$ has positive measure.
\end{proof}

\subsection{Excursion: global minimisers on $\vz T_L$}
\label{subsec:ex_on_TL}

By very similar arguments one may prove the corresponding statement for
functions on the torus~$\vz T_L$, thus extending the characterisation of~\cite{ChoksiRen05} to all global minimisers.
\begin{theorem}
\label{th:exist_periodic_1d}
Let $L>0$, $d_{uv} > 0$, and fix a mass $M>0$, with $M<L/2$.
\begin{enumerate}
\item There exists a global minimiser $(u, v)$ of $F_1$ on the torus $\vz T_L$ under constrained mass $M$ for which $\supp(u+v)$ is connected.\label{item:torusconnmin}
\item This global minimiser is \emph{non}-unique (apart from translation and
mirroring) if and only if\label{item:torusnonunique}
  \begin{enumerate}
    \item the energy of this configuration is equal to the energy of another
configuration $(\bar u, \bar v)$ for which $\supp(\bar u+\bar v)$ is connected,
or \label{item:sameenergy}
    \item one of the following two conditions is satisfied:
    \label{th:periodic:cond}
    \begin{enumerate}
      \item $d_{u0}=0$ and there exists a global minimiser with an internal U-block or;\label{item:PER:condunique1}
      \item $d_{v0}=0$ and there exists a global minimiser with an internal V-block.
            \label{item:PER:condunique2}
    \end{enumerate}
  \end{enumerate}
\end{enumerate}

\end{theorem}

\begin{proof}
The proof follows the same lines as in the case of $\R$, Theorem~\ref
{th:exist_real_line}. We will point out the differences between the two cases.

Suppose $(u, v) \in K_1$ is a global minimiser such that $\vz{T}_L\setminus
\supp(u+v)$ has at least two connected components, which, by translating $u$ and
$v$, we can assume to be $(x_1,x_2)\subset [0,L)$ and $(x_3,L)$ with $x_3>x_2$.
Let $\phi$ be the associated potential; since $u$ and $v$ vanish on $(x_1,x_2)$
and $(x_3,L)$, $\phi$ is linear on these two intervals. Let $\phi'(x) = \phi'_
{12}$ for $x \in [x_1, x_2]$ and $\phi'(x) = \phi'_{3L}$ for $x \in [x_3, L]$.
By possibly exchanging roles we can assume that $|\phi'_{12}| \geq |\phi'_{3L}|
$.

Constructing for some $0<a<x_2-x_1$ the same translated functions $\bar u$, $\bar v$, and $\widetilde \phi$ as given in (\ref{def:overline_u}--\ref{def:widetilde_phi}), we have the
analogous inequality
\begin{align}
\notag
\int_0^L {\widetilde\phi'^2}
 &= \int_0^{x_1}{\phi'}^2 + \int_{x_1+a}^{x_3} {\phi'}^2 +
     {\phi'}_{3L}^2 (L-x_3+a) \\
     \label{ineq:PER:phi1phi3}
 &\leq \int_0^{x_1}{\phi'}^2 + \int_{x_1+a}^{x_3} {\phi'}^2 +
     a{\phi'}_{12}^2 + {\phi'}_{3L}^2(L-x_3)\\
 &= \int_0^L {\phi'}^2.
 \notag
\end{align}
Although $\widetilde\phi$ satisfies ${\widetilde \phi}'' = \bar u - \bar v$
on $(0,L)$, the function $\widetilde \phi$ can in general not be extended
periodically, i.e. $\widetilde\phi(0)\not=\widetilde\phi(L)$. To correct
this we define
\[
\bar \phi(x) := \widetilde\phi(x) - \frac xL (\widetilde\phi(L)-\widetilde\phi(0)),
\]
so that the function $\bar \phi$ solves ${\widetilde \phi}'' = \bar u - \bar v$
on $(0,L)$, is continuously differentiable on $(0, L)$, and satisfies $\bar \phi
(0) = \bar \phi(L)$. From
\[
\bar\phi'(L)-\bar\phi'(0) = \int_0^L \bar \phi'' = \int_0^L (v-u) = 0,
\]
we conclude $\bar\phi'(0) = \bar\phi'(L)$, so that $\bar \phi$ is the Poisson
potential on $\vz T_L$ associated with $\bar u$ and $\bar v$.
In addition,
\begin{align}
\notag
\int_0^L \bar \phi'^2 &= \int_0^L {\widetilde\phi}'^2
  - \frac2L(\widetilde\phi(L)-\widetilde\phi(0)) \int_0^L \widetilde\phi'
  + \frac1L(\widetilde\phi(L)-\widetilde\phi(0))^2 \\
  &= \int_0^L {\phi'}^2 -  \frac1L(\widetilde\phi(L)-\widetilde\phi(0))^2\notag\\
  &\leq \int_0^L {\phi'}^2.
  \label{ineq:averaging}
\end{align}

From these two inequalities it follows as in the proof of Theorem~\ref
{th:exist_real_line} that $F_1(\bar u,\bar v)\leq F_1(u,v)$, so that existence
of any global minimiser again implies the existence of a connected global
minimiser.

We now turn to the discussion of the necessary and sufficient conditions for
non-uniqueness. Again we use the fact that inequalities are saturated to deduce
necessary conditions; in this case, however, there is an additional inequality
in~\pref{ineq:averaging}. The reasoning proceeds in two steps.

\textbf{\boldmath Step 1: Take $a<x_2-x_1$.}
When $a<x_2-x_1$ no interfaces are created, annihilated or changed, and we only
need to consider the inequalities in~\pref{ineq:PER:phi1phi3} and~\pref
{ineq:averaging}. Since these are saturated, the following conditions hold:
\begin{enumerate}
\item $|\phi'_{12}| = |\phi'_{3L}|$, and \label{item:saturation1}
\item $\widetilde\phi(L)=\widetilde\phi(0)$.\label{item:saturation2}
\end{enumerate}
We first calculate
\begin{align*}
\widetilde\phi(L)-\widetilde\phi(0) &= \phi(x_3)-\phi(x_1+a)+\phi(x_1)
  + \phi'_{3L}(L-x_3+a) - \phi(0) \\
  &= -\phi(x_1+a)+\phi(x_1) + a \phi'_{3L} \\
  &= a (\phi'_{3L} - \phi'_{12}).
\end{align*}
By condition~\ref{item:saturation2} above we have $\phi'_{12}=\phi'_{3L}$, which
is also compatible with condition~\ref{item:saturation1}.

We now claim that $\phi'_{12} = \phi'_{3L} = 0$. Suppose not, say (for
concreteness) $\phi'_{12}=\phi'_{3L}>0$, then $\phi(x_1) < \phi(x_2)$ and $\phi
(x_3) < \phi(L)$. Since for a stationary point the potential $\phi$ has the same
value at all U-0 type transitions and the same value for all V-0 type transitions, the two
transitions at $x_1$ and at $x_2$ are of different type, thus one is a U-0 type
transition and the other a V-0 type transition. The same is true for $x_3$ and $L$
(or $0$); and the transitions at $x_1$ and $x_3$ are the same. Therefore $\phi
(x_1) =\phi(x_3)$.

For any fixed $a$ in the interval $(0,x_2-x_1)$, however, we have now constructed a
second global minimiser $(\bar u,\bar v)$---and therefore a second stationary point---for which $\phi(x_1) \neq \phi(x_3-a)$, since $\bar \phi(x_1) = \widetilde\phi(x_1) = \phi
(x_1)$ and
\begin{align*}
\bar\phi(x_3-a) =\widetilde\phi(x_3-a) &= \phi(x_3) - \phi(x_1+a)+\phi(x_1) \\
&= \phi(x_1) - a\phi'_{12}\\
&< \phi(x_1).
\end{align*}
Since the interfaces of $(\bar u, \bar v)$ at $x_1$ and $x_3-a$ are of the same
type, this contradicts the stationarity of this second minimiser, and we
conclude that $\phi'_{12}= \phi'_{3L} = 0$. Note that since the intervals $
(x_1,x_2)$ and $(x_3,L)$ were chosen as arbitrary connected components of $\vz
{T}_L\setminus\supp(u+v)$, this implies that $\phi'$ vanishes on the whole of $
\vz{T}_L\setminus\supp(u+v)$.

\textbf{\boldmath Step 2: Take $a=x_2-x_1$.}
Non-uniqueness in this case implies that also the interfacial energy remains the
same in the construction of $(\bar u, \bar v)$. As in the case of $\R$, the
interfaces at $x_1$ and $x_2$ that are joined together in the construction of $
(\bar u, \bar v)$ are of the same type, \emph{i.e.} either both U-0 type or both V-0 type
transitions. The fact that $\phi$ is constant on 0-blocks is used in this
argument. We conclude that either $d_{u0}=0$ or $d_{v0}=0$, and that a connected
global minimiser exists with at least one internal U-block (if $d_{u0}=0$) or at
least one internal V-block (if $d_{v0}=0$). This proves the necessity of
condition~\ref{th:periodic:cond}.

The sufficiency of condition~\ref{th:periodic:cond} follows by splitting one of
the internal blocks, as in the case of~$\R$. Apart from the simplifying fact that the torus is bounded, the proof of existence of a global
minimiser is identical to the case of $\R$.
\end{proof}

\begin{remark}
Note that the proof of existence of a global minimiser generalises straightforwardly to the higher dimensional case of the torus $\vz{T}_L^N$, because the torus is bounded. On the unbounded domain $\R^N$, $N\geq 2$, the above proof does not suffice.
\end{remark}

\subsection{Explicit values and a lower bound}
\label{sec:lower_bnd_1d}

We now focus again on functions on $\vz{R}$. The results of the previous sections allow us to calculate global minima of the
energy $F_1(u,v)$ as a function of the mass $M=\int u$. Two important special cases are the
monolayer and the bilayer.

A \emph{monolayer} consists of a single U- and a single V-block, of equal width
$m$, where $m$ is the mass of $u$ or $v$, i.e. positioning the block around the
origin for convenience,
\[
u(x) = \chi^{}_{(-m, 0)}
\quad\text{and}\quad
v(x) = \chi^{}_{(0, m)},
\]
where $\chi^{}_A$ is the characteristic function of the set $A$. We then find for the derivative
of the Poisson potential
\[
\phi'(x) = \begin{cases}
0 & \text{for $x< -m$}\\
 |x| -m & \text{for $-m<x<m$}\\
0 & \text{for $x>m$},
\end{cases}
\]
The total energy then becomes
\[
\text{monolayer of mass $M=m$:}
\qquad F_1 = 2(c_0+c_u+c_v)
  + \frac23 m^3.
\]
Note the definition of mass: a monolayer of mass $M$ means that $\int u = \int v
= M$, and therefore that the `total' mass of the monolayer $\int (u+v)$ equals
$2M$. In this case the mass $M$ of the monolayer equals the width $m$ of each of the blocks.

A \emph{bilayer} consists of two monolayers joined back-to-back. It comes in two
varieties, as UVU and as VUV. For a UVU bilayer of mass $M=2m$, given by
\[
u(x) = \chi^{}_{(-2m, -m) \cup (m, 2m)}
\quad\text{and}\quad
v(x) = \chi^{}_{(-m, m)},
\]
the derivative of the Poisson potential is
\[
\phi'(x) = \begin{cases}
0 & \text{for $x< -2m$}\\
-2m - x & \text{for $-2m<x<-m$}\\
x & \text{for $-m<x<m$}\\
2m - x  & \text{for $m<x<2m$}\\
0& \text{for $x>2m$}.
\end{cases}
\]
The energy has the value
\[
\text{UVU bilayer of mass $M=2m$}: \qquad F_1 = 2c_0+4c_u + 2c_v
  + \frac43 m^3,
\]
For a VUV bilayer the situation is of course analogous:
\[
\text{VUV bilayer of mass $M=2m$}: \qquad F_1 = 2c_0+2c_u + 4c_v
  + \frac43 m^3.
\]

Similarly, \emph{$n$-layered structures} consisting of $n$ monolayers back-to-back, have energy
\begin{align}
\label{explicit_energy_multilayer_1}
 \text{VUVU\ldots V $n$-monolayer with mass $M=n m$}: \qquad
  F_1 &= 2c_0+nc_u + (n+2)c_v + \frac{2n}3 m^3 \\
  &= 2d_{v0} + n d_{uv} + \frac{2n}3 m^3,
\label{explicit_energy_multilayer_1a}\\
\label{explicit_energy_multilayer_2}
 \text{VUVU\ldots U $n$-monolayer with mass $M=n m$}: \qquad
  F_1 &= d_{u0} + d_{v0} + n d_{uv} + \frac{2n}3 m^3,\\
\label{explicit_energy_multilayer_3}
 \text{UVUV\ldots U $n$-monolayer with mass $M=n m$}: \qquad
  F_1 &= 2d_{u0} + n d_{uv} + \frac{2n}3 m^3.
\end{align}
Note that for a VUVU\ldots V $n$-monolayer or UVUV\ldots U $n$-monolayer the value of $n$ is even, while for a VUVU\ldots U $n$-monolayer it is odd. Furthermore $m$ is the U-mass in one monolayer, thus $m$ is the width of the outer blocks, from which we see that the width of the inner blocks is $2m$.
By collecting these results we find:
\begin{theorem} \label{th:lower_bound_1d}
Let $N=1$. For any structure of mass $M$,
\begin{equation}
\label{bound:lower_bound_1d}
F_1 \geq 2(c_0+\min(c_u,c_v)) + \left(\frac92\right)^{1/3}d_{uv}^{2/3} M.
\end{equation}
In the limit of large mass,
\begin{equation}
\label{limit:Mtoinfty}
\lim_{M\to\infty} \inf\left\{ \frac{F_1(u,v)}M: (u,v)\in K_1,\ \int_\R u = M\right\} = \left(\frac92\right)^{1/3}d_{uv}^{2/3}.
\end{equation}
\end{theorem}
\begin{proof}
If $d_{uv}=0$, then the first statement is easily checked and the second follows from the example of Section~\ref{subsec:d_12}. We continue under the assumption that $d_{uv}>0$.

Let $(u, v)$ be a global minimiser with connected $\supp(u+v)$, which exists
according to Theorem~\ref{th:exist_real_line}. Note that for all three cases of structures (VUVU\dots V, UVUV\dots V, and UVUV\dots U) the interfacial terms are bounded from below by $2(c_0 + \min(c_u,c_v))$, so that
\[
F_1 \geq 2(c_0 + \min(c_u,c_v)) + nd_{uv}+ \frac2{3n^2} M^3.
\]
Minimising this with respect to $n$ gives the desired lower
bound. The particular value of $n$ for which the lower bound is achieved,
\[
n_0(M)^3:= \frac43 \,\frac{M^3}{d_{uv}},
\]
will be useful below.

To prove the second part of the theorem, we note that (\ref{explicit_energy_multilayer_1a}-\ref{explicit_energy_multilayer_3}) imply the upper bound
\begin{equation}
\label{ineq:upperboud_F1_M}
\inf \left\{\frac{F_1(u,v)}M: (u,v)\in K_1, \int u = M\right \} \leq
  \frac2M\max\{d_{u0},d_{v0}\}
  + \inf_{n\in\N} \left\{ \frac nM d_{uv} + \frac{2}{3} \left(\frac Mn\right)^2 \right\}.
\end{equation}
Choosing the largest integer smaller or equal to $n_0(M)$ as particular value of $n$,
\[
n(M) := \left\lfloor {}^3\sqrt{\frac43\, \frac{M^3}{d_{uv}}}\right\rfloor = \lfloor n_0(M) \rfloor,
\]
we have $n_0(M)-1 < n(M)\leq n_0(M)$. In the limit $M\to\infty$ the quotient $n(M)/M$ therefore converges to $(4/3\,d_{uv})^{1/3}$; with this convergence the inequality~\pref{ineq:upperboud_F1_M} implies~\pref{limit:Mtoinfty}.
\end{proof}

\begin{figure}[htb]
\centering
    \psfrag{1}{$1$}
    \psfrag{2}{$2$}
    \psfrag{3}{$3$}
    \psfrag{4}{$4$}
    \psfrag{5}{$5$}
    \psfrag{6}{$6$}
    \psfrag{8}{$8$}
    \psfrag{a}{$1.17$}
    \psfrag{b}{LB}
    \psfrag{x}{$M$}
    \psfrag{y}{$\frac{F_1}{M}$}
    \includegraphics[width=110mm]{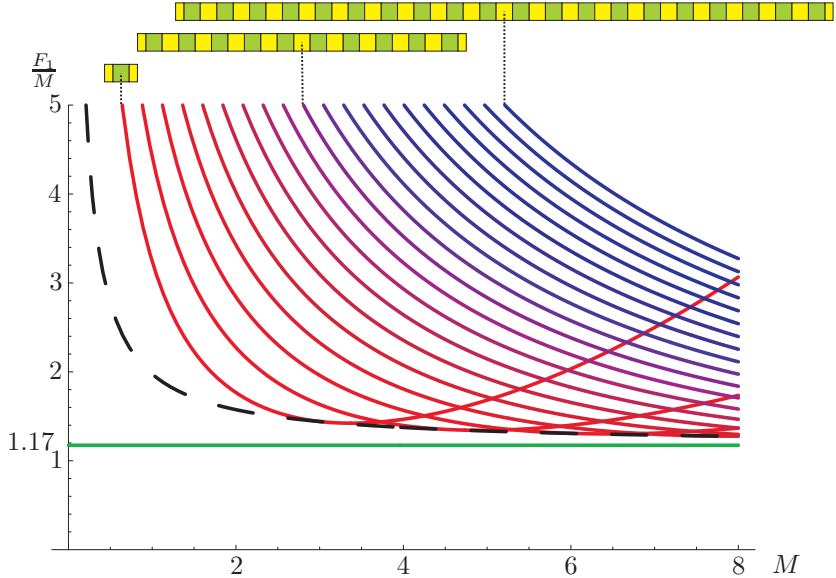}\\
    \caption{Energy per unit mass for the one-dimensional case, according to the
calculations in Section~\ref{sec:lower_bnd_1d}. $M$ is the total U-mass; for the parameters the values $d_{u0} = 1, d_{uv} = 0.6$ and $d_
{v0} = 0.4$ are chosen. All the graphs belong to a VUVU\dots V $n$-monolayer
structure, where $n/2$ increases from $1$ (left) to $20$ (right) with step size $1$. Also drawn are the (dashed) lower bound LB~\pref{bound:lower_bound_1d}, and the
asymptote $2^{-\frac13} 3^{\frac23} d_{uv}^{\frac23} \approx 1.17446$.} \label{fig:blocks_asymp}
\end{figure}

In Figure~\ref{fig:blocks_asymp} the graphs depicting the energy per mass for
VUVU\dots V configurations consisting of different numbers of monolayers are
shown, for some specific parameter values. The lower bound from Theorem~\ref
{th:lower_bound_1d} is indicated as well.

\begin{remark}\label{rem:convergenceofwidth}
Minimising $F_1/M$ from (\ref{explicit_energy_multilayer_1}--\ref{explicit_energy_multilayer_3}) with respect to $m$, we find as minimising value for $m$,
\[
m_0^3(n) := \frac{3(k_1 d_{u0}+k_2 d_{v0}+n d_{uv})}{4n},
\]
where, depending on the configuration $k_1=0, k_2=2$ (\ref{explicit_energy_multilayer_1a}), $k_1=k_2=1$ (\ref{explicit_energy_multilayer_2}) or $k_1=2, k_2=0$ (\ref{explicit_energy_multilayer_3}). In all three cases we find that in the limit $n\to\infty$, or equivalently (for $m$ fixed) $M\to\infty$, the width of the inner blocks converges to
\[
2 \lim_{n\to\infty} m_0(n) = 6^{1/3} d_{uv}^{1/3}.
\]
Note that in \cite{Mueller93} and \cite{RenWei03a} it is found that one-dimensional minimisers for the functionals under consideration in those papers are periodic with period $\sim (\text{surface tension})^{1/3}$. (In these diffuse interface functionals the surface tension coefficients are given by integrating the square root of the potential.)
\end{remark}

\section{Higher dimensions}\label{sec:scaling}

In this section we derive bounds on energy of minimisers in terms of the mass $M$. The first result, Theorem~\ref{lem:umassineqs}, shows that the minimal energy has a lower bound that scales linearly in mass in the limit $M\to\infty$. This is an extension of the lower bound~\pref{bound:lower_bound_1d} in one dimension, but with a smaller constant.

A simple argument immediately gives an upper bound on the minimal energy at given mass: fixing any structure of unit mass, a candidate structure at mass $M\in \N$ can be obtained by distributing $M$ copies of the unit-mass structure over $\R^N$. The energy of the resulting structure equals $M$ times the energy of the unit-mass structure. This construction can be extended to non-integer mass $M$ by spatially stretching a structure of integer mass close to $M$. In the limit $M\to\infty$ the resulting perturbation of the energy is small.

In Sections~\ref{sec:upper_bound} and~\ref{subsec:examples} we therefore provide tighter upper bounds, by constructing $N$-dimensional structures out of near-optimal $k$-dimensional ones, with $k<N$.

\subsection{Lower bound}
\label{subsec:lower_bound}
For this section we pick a function $\kappa\in C_c^\infty(\R^N)$, non-negative and radially symmetric, such that
\[
\int_{\vz{R}^N} \kappa = 1.
\]

For $\epsilon > 0$ we now define
\[
\kappa_{\epsilon}(x) := \frac{1}{\epsilon^N} \kappa(x/\epsilon).
\]
Note that $\int_{\vz{R}^N}\kappa_{\epsilon} = 1$ for all $\epsilon$.
In the following we will use the constant $A_N$, defined as
\[
A_N :=  \dashint_{S^{N-1}} |e \cdot w| \,d\mathcal{H}^{N-1}(w),
\]
where $S^{N-1}$ is the $(N-1)$-dimensional unit sphere and $e$ is any element of $S^{N-1}$. This definition is independent of the choice of $e$, because the integration is over all of $S^{N-1}$.

The central result is an interpolation inequality between the $BV$-seminorm and
$H^{-1}$. In spirit, and in its application, it is similar to the Lemma 2.1 of~\cite{Choksi01}. The proof is different, however, and uses an argument of~\cite{KohnOtto02}, in combination with the characterisation of $BV$ by~\cite{Davila02}.
\begin{theorem}\label{lem:umassineqs}
Let $d_{uv}\neq 0$. For all $(u,v)\in K_1$,
\begin{equation}\label{eq:inequmass1}
\int_{\R^N} u \leq C_1(\kappa, N) \|u - v
\|_{H^{-1}(\R^N)}^{\frac{2}{3}} \left( \int_{\R^N} |\nabla u|
\right)^{\frac{2}{3}},
\end{equation}
where $C_1(\kappa, N) > 0$ is given by
\[
C_1(\kappa, N) := 2^{\frac43} A_N^{\frac23} \left(\int_{\vz{R}^N} |\nabla
\kappa|\right)^{\frac23} \left(\int_{\vz{R}^N} |y| \kappa(y)\,dy\right)^
{\frac23}.
\]
The inequality~\pref{eq:inequmass1} also holds with $u$ and $v$ interchanged.
Furthermore,
\begin{equation}
\label{est:lower_bound_nD}
F_1(u,v) \geq C_2(\kappa,N) \int_{\R^N} u,
\end{equation}
where
\[
C_2(\kappa,N) := \frac32 C_1(\kappa,N)^{-1} \left( c_u^{3/2}+c_v^{3/2}\right).
\]
\end{theorem}

\begin{proof}
If $\int_{\R^N}u=0$ the statements are trivially true. In what follows we assume $\int_{\R^N}u>0$.

First note that $\kappa_\epsilon\ast u \in H_0^1\left(\R^N\right)$.
From $uv=0$ it follows that $v\leq 1 -u$, so that
\[
\int_{\R^N} (u - v) \kappa_{\epsilon} \ast u \geq \int_{\R^N} \left(2 u - 1\right)
\kappa_{\epsilon} \ast u = 2 \int_{\R^N} u \kappa_{\epsilon} \ast u -
 \int_{\R^N} u,
\]
Writing
\begin{align*}
2 \int_{\R^N} u \kappa_{\epsilon} \ast u
 &= 2\int_{\R^N}\int_{\R^N} u(x)u(y)\kappa_\e(x-y)\, dxdy \\
 &= -\int_{\R^N}\int_{\R^N} (u(x)-u(y))^2 \kappa_\e(x-y) \, dxdy
    + 2\int_{\R^N} u^2\\
 &= - \int_{\R^N}\int_{\R^N} |u(x)-u(y)| \kappa_\e(x-y) \, dxdy
    + 2 \int_{\R^N} u,
\end{align*}
we have
\[
\int_{\R^N} u \leq \int_{\R^N} (u - v) \kappa_{\epsilon} \ast u
  + \int_{\R^N}\int_{\R^N} |u(x)-u(y)| \kappa_\e(x-y) \, dxdy.
\]
The first term on the right-hand side is estimated by combining the definition
of the $H^{-1}$-norm,
\[
\int_{\R^N} (u - v) \kappa_{\epsilon} \ast u \leq \|u -v \|_{H^{-1}} \|\nabla
\kappa_{\epsilon} \ast u \|_{L^2},
\]
with the estimate (Young's inequality~\cite[Theorem~4.30]{Adams75})
\[
\| \nabla \kappa_{\epsilon} \ast u \|_{L^2}
  \leq \| u \|_{L^2} \int_{\vz{R}^N} |\nabla \kappa_{\epsilon}|
  = \|u\|_{L^1}^{\frac12} \int_{\vz{R}^N} |\nabla \kappa_{\epsilon}|
  = \epsilon^{-1}  \|u\|_{L^1}^{\frac12} \int_{\vz{R}^N} |\nabla \kappa|.
\]

For the second term we use a density argument as in \cite[proof of Lemma~3]{Davila02} to find
\begin{align*}
\lefteqn{ \int_{\R^N}\int_{\R^N} |u(x)-u(y)| \kappa_\e(x-y) \, dxdy \leq {}} \qquad &
\\
&\leq
\e \int_{\R^N} \int_{\vz{R}^N} \int_0^1 \left|\nabla u(t y
+ (1 - t) x) \frac{(y - x)}{|y - x|} \right| \frac{|y -
x|}{\epsilon}
\kappa_{\epsilon}(x - y) \, dt \, dy \, dx\\
&= \e\int_{\vz{R}^N} \int_0^1\int_{\R^N}\left|\nabla u(x
+ t h) \cdot \frac{h}{|h|} \right| \frac{|h|}{\epsilon}
\kappa_{\epsilon}(h) \, dx\, dt \, dh\\
&= \e \int_{\vz{R}^N} \int_{\R^N} \left|\nabla u(z) \cdot
\frac{h}{|h|} \right| \frac{|h|}{\epsilon} \kappa_{\epsilon}(h) \,
dz \, dh\\
&= \e \int_0^{\infty} \int_{\R^N} \int_{S^{N-1}} |\nabla
u(z) \cdot w|\, d\mathcal{H}^{N-1}(w)\,  r^{N-1} \frac r{\epsilon}
\kappa_{\epsilon}(r)
\, dz \, dr\\
&= \e A_N \, \mathcal{H}^{N-1}\left(S^{N-1}\right)\int_{\R^N} |\nabla u(z)| \,
dz\,
\int_0^{\infty} \frac{r^N}{\epsilon} \kappa_{\epsilon}(r) \, dr\\
&= \e A_N \int_{\vz{R}^N} |y|
\kappa(y) \, dy \int_{\R^N} |\nabla u(z)| \, dz.
\end{align*}
The first equality follows after substituting $y = x+h$, while
the substitution $x =z- th$ leads to the second equality.

Collecting the parts we find the estimate
\[
\int_{\R^N} u \leq \epsilon^{-1}  \|u - v\|_{H^{-1}}
   \left(\int_{\R^N} u\right)^{\frac12} \int_{\vz{R}^N}|\nabla \kappa|
   + \epsilon C_0(\kappa, N) \int_{\R^N} |\nabla u|,
\]
where
\[
C_0(\kappa, N) := A_N \int_{\vz{R}^N} |y| \kappa(y) \, dy.
\]
Minimising the right hand side with respect to $\epsilon$ we find
\[
\int_{\R^N} u \leq 2  \left[C_0(\kappa, N) \int_{\vz{R}^N} |\nabla \kappa| \left
(\int_{\R^N} u\right)^{\frac12} \|u-v\|_{H^{-1}} \int_{\R^N} |\nabla u|. \right]^
{\frac12}
\]
Dividing both sides by $\left(\int_{\R^N} u\right)^{\frac14}$ and then raising them
both to the power $4/3$ gives the first statement of the theorem. Since we have used no property that distinguishes $u$ from $v$, we can apply the same argument with $u$ and $v$ interchanged.

To prove the inequality~\pref{est:lower_bound_nD}, we remark that from~\pref{eq:inequmass1} and Young's Inequality we obtain, for any $\alpha, \beta>0$,
\begin{align*}
C_1^{-1} \int u &\leq \frac {2\alpha}3 \int |\nabla u|
  + \frac1{3\alpha^2} \|u-v\|_{H^{-1}}^2,\\
C_1^{-1} \int u &\leq \frac {2\beta}3 \int |\nabla v|
  + \frac1{3\beta^2} \|u-v\|_{H^{-1}}^2.
\end{align*}
By choosing
\[
\alpha := c_u^{1/3}
\qquad\text{and}\qquad
\beta := c_v^{1/3},
\]
and then adding the two inequalities with weights $\alpha^2$ and $\beta^2$ respectively, estimate~\pref{est:lower_bound_nD} follows.
\end{proof}
Note from the proof above that estimate~(\ref{est:lower_bound_nD}) is not sharp if $\int_{\R^N}u>0$.
\begin{remark}
Inequality~(\ref{est:lower_bound_nD}) does not hold in the case where $d_{uv} = 0$. The same sequence $(u_n,v_n)$ that was introduced in Section~\ref{subsec:d_12} demonstrates this fact, since  $F_1(u_n,v_n)\to 2$ while $\int u_n \to\infty$.
\end{remark}

\subsection{Upper bound}
\label{sec:upper_bound}

We next show that the one-dimensional upper bound~\pref{limit:Mtoinfty} (or~\pref{est:upper_bound_1d}) also
holds in higher dimensions, as a consequence of the more general statement
below. Theorem~\ref{th:cutoff} formalises the intuitive idea that extending a
one-dimensional minimiser in the other directions, and then cutting off the
resulting planar structure at some large distance, should result in an $N$-
dimensional structure whose energy-to-mass ratio is close to that of the
original one-dimensional structure. We formulate the result for $k$-dimensional
structures that are embedded in $N$ dimensions.

Let $1\leq k\leq N-1$, and let us write $K_{1,k}$ for the admissible set $K_1$
on $\R^k$. Let $(\overline u , \overline v)$ be
\begin{itemize}
\item any element of $K_{1,k}$, when $k\geq 3$; or
\item any element of $K_{1,k}$ with $\int_{\R^k} x(\overline u(x)-\overline v
(x))\, dx= 0$, when $k\in{1,2}$.
\end{itemize}
(We explain this restriction in Remark~\ref{rem:firstmoment}). Split vectors $x
\in \R^N$ into two parts, $x=(\xi, \eta)\in\R^k\times \R^{N-k}$, and define a
cutoff function $\chi_a:\R^{N-k}\to[0,1]$ by
\[
\chi_a(\eta) := \chi(|\eta|-a),
\]
where $\chi:\R\to[0,1]$ is fixed, smooth, and satisfies $\chi(x)=1$ for $x\leq0
$, $\chi(x) = 0$ for $x\geq1$. We will compare the energy values of the $k$-
dimensional structure $(\overline u,\overline v)$ with those of the $N$-
dimensional structure
\begin{equation}\label{eq:extension}
(u,v)(x) := (\overline u,\overline v)(\xi)\chi_a(\eta).
\end{equation}
Note that this $(u,v)$ is an element of $K_{1,N}$, the admissible set $K_1$ on $
\R^N$.

A note on notation: $\omega_d$ will denote the $d$-dimensional Lebesgue measure of the $d$-dimensional unit ball.

\begin{theorem}
\label{th:cutoff}
Fix $(\overline u,\overline v)$ as given above. Then, for $(u,v)$ as defined in (\ref{eq:extension}),
\[
\frac{F_1(u,v)}{\int_{\R^N} u}
  = \frac{F_1(\overline u,\overline v)}{\int_{\R^k} \overline u} + O(1/a)
\qquad\text{as }a \to \infty.
\]
\end{theorem}

\begin{proof}
We first estimate the interfacial terms as follows:
\begin{align}
\notag
\int_{\R^N} |\nabla u|
  &= \int_{\R^k} \int_{\R^{N-k}}|\nabla\overline u(\xi)|\, \chi^{}_a(\eta)
       \, d\eta d\xi
     + \int_{\R^k} \int_{\R^{N-k}} \overline u(\xi)|\nabla \chi^{}_a(\eta)|
       \, d\eta d\xi \\
  &\begin{cases}
     \;\leq\; \ds\omega_{N-k} (a+1)^{N-k} \int_{\R^k} |\nabla \overline u|
            + (N-k)\omega_{N-k} (a+1)^{N-k-1} \|\chi'\|_\infty \int_{\R^k}
\overline u,\\
     \;\geq\; \ds\omega_{N-k} a^{N-k} \int_{\R^k} |\nabla \overline u|,
   \end{cases}
   \label{ineq:0u}
\end{align}
and therefore
\[
\int_{\R^N} |\nabla u| = (1+O(1/a))\omega_{N-k} a^{N-k} \int_{\R^k} |\nabla
\overline u|\qquad\text{as }a \to \infty.
\]
Similarly,
\begin{align}
\label{ineq:0v}
\int_{\R^N} |\nabla v| &= (1+O(1/a))\omega_{N-k} a^{N-k}\int_{\R^k} |\nabla
\overline v| \qquad \text{and}\\
\int_{\R^N} |\nabla (u+v)| &= (1+O(1/a))\omega_{N-k} a^{N-k}\int_{\R^k} |\nabla
(\overline u+\overline v)|
\label{ineq:0uv}
\end{align}
The estimate of the $H^{-1}$-norm is formulated in Theorem~\ref{thm:HmO-estimate}. The result now follows by combining the estimates (\ref{ineq:0u}--\ref{ineq:periodic_and_not}) and remarking that the mass of $(u,v)$ is given by
\[
\int_{\R^N} u = \int_{\R^N} \overline u(\xi)\chi_a(\eta)
= (1+O(1/a))\omega_{N-k}a^{N-k}\int_{\R^k} u(\xi).
\]
\end{proof}

\begin{theorem}
\label{thm:HmO-estimate}
Under the conditions above there exists a constant $C=C(k,N)$ such that for all
$a>0$,
\begin{equation}
\label{ineq:periodic_and_not}
\Bigl|\|u-v\|^2_{H^{-1}(\R^N)}
  - \omega_{N-k} a^{N-k}\|\overline u-\overline v\|^2_{H^{-1}(\R^k)}\Bigr|
\leq Ca^{N-k-1}\int_{\R^k} \bigl[|\nabla \overline\phi|^2+\overline\phi^2\bigr].
\end{equation}
Here $\overline\phi$ is a $k$-dimensional Poisson potential associated with $
(\overline u, \overline v)$.
\end{theorem}
\begin{remark}
\label{rem:firstmoment}
The restriction of vanishing first moments for $k=1,2$ follows directly from the
requirement that $\int_{\R^k}\overline\phi^2$ can be chosen finite in~\pref
{ineq:periodic_and_not}. Since the integral of $\overline u-\overline v$
vanishes the potential $\overline \phi := G*(\overline u - \overline v)$ decays
to zero at least as fast as $|\xi|^{1-k}$, as can be seen from the multipole expansion of $\bar \phi$ (see \cite{HohlfeldKingDruedingSandri93}). For dimensions $k\geq 3$ it follows
that $\int \overline \phi^2$ is finite; but for $k=1,2$ a higher decay rate is
necessary, which we provide by requiring an additional vanishing moment. The
case $k=1$ is special: the vanishing of the zero and first moments implies that
$\overline\phi := G*(\overline u - \overline v)$ is zero in a neighbourhood of
infinity.
\end{remark}
\begin{proof}[Proof of Theorem~\ref{thm:HmO-estimate}]
The Poisson potential $\phi$ associated with $(u,v)$ satisfies
\[
-\Delta \phi(x) = u(x)-v(x) = (\overline u(\xi)-\overline v(\xi))\chi_a(\eta)
\qquad\text{for }x=(\xi,\eta)\in\R^N.
\]
Similarly, the $k$-dimensional potential $\overline \phi$ associated with $
(\overline u, \overline v)$ satisfies
\[
-\Delta_\xi\overline\phi(\xi) = \overline u(\xi)-\overline v(\xi) \qquad\text{for }
\xi\in\R^k.
\]
We write $\nabla_\xi$ for the part of the gradient that operates on $\xi$, that
is
$(\partial_{x_1},\partial_{x_2},\dots,\partial_{x_k},0,\dots,0)$, and we use a
similar notation for the other part of the gradient $\nabla_\eta$ and the
partial Laplacians $\Delta_\xi$ and $\Delta_\eta$. Remarking that
\begin{align*}
\int_{\R^N} \chi_a(\eta) \nabla \phi(x)\cdot \nabla \overline \phi(\xi)\, dx
  &= \int_{\R^N} \chi_a(\eta) \nabla_{\xi} \phi (x) \cdot \nabla_\xi \overline
\phi(\xi)\, dx
  = -\int_{\R^N}\chi_a(\eta) \phi(x) \Delta_\xi \overline\phi(\xi) \, dx\\
  &= \int_{\R^N} \chi_a(\eta) \phi(x) (\overline u(\xi)-\overline v(\xi))\, dx
  = -\int_{\R^N}\phi(x)\Delta \phi(x) \, dx\\
  &= \int_{\R^N} |\nabla \phi(x)|^2 \, dx,
\end{align*}
we calculate
\begin{align}
\notag
\int_{\R^N} |\nabla\phi - \chi_a \nabla \overline \phi|^2
  &= \int_{\R^N} |\nabla \phi|^2 - 2\int_{\R^N} \chi_a\nabla\phi\nabla\overline
\phi
      + \int_{\R^N} \chi_a^2 |\nabla\overline\phi|^2\\
  &= - \int_{\R^N} |\nabla \phi|^2 + \int_{\R^N} \chi_a^2 |\nabla\overline\phi|
^2.
  \label{eq:discrepancy}
\end{align}
One inequality relating the two norms can be deduced directly:
\[
\|u-v\|_{H^{-1}(\R^N)}^2 = \int_{\R^N} |\nabla \phi|^2
  \leq \int_{\R^N} \chi_a^2 |\nabla\overline\phi|^2
  \leq \omega_{N-k} (a+1)^{N-k} \|\overline u - \overline v\|^2_{H^{-1}(\R^k)}.
\]

For the opposite inequality we set
\[
\psi(x) := \phi(x) - \overline\phi(\xi)\chi_a(\eta),
\]
and rewrite
\begin{align*}
\int_{\R^N} |\nabla\phi - \chi_a \nabla \overline \phi|^2
  &= \int_{\R^N} |\nabla\psi|^2
    + 2\int_{\R^N} \overline \phi \nabla_\eta\psi\nabla_\eta\chi_a
    + \int_{\R^N}\overline \phi^2 |\nabla\chi_a|^2\\
  &= \int_{\R^N} |\nabla\psi|^2
    - 2\int_{\R^N} \psi\overline \phi\Delta_\eta\chi_a
    + \int_{\R^N}\overline \phi^2 |\nabla\chi_a|^2 \\
  &=: I(\psi).
\end{align*}
Since
\[
-\Delta\psi = -\Delta \phi + \chi_a\Delta_\xi\overline\phi + \overline \phi
\Delta_\eta\chi_a
  = \chi_a (\overline u-\overline v) - \chi_a(\overline u - \overline v)
     + \overline\phi\Delta_\eta\chi_a
  = \overline\phi\Delta_\eta\chi_a,
\]
the function $\psi$ is the global minimiser of $I$, which is convex as functional on $\psi$. Therefore, setting
$\psi_0(x) := \overline\phi(\xi) |\nabla\chi_a(\eta)|^2$,
\begin{align*}
I(\psi) \leq I(\psi_0)
  &= \int_{\R^N} \left[ |\nabla\overline\phi|^2 |\nabla\chi_a|^4
      + 4\overline\phi^2 |D^2\chi_a\cdot\nabla\chi_a|^2
      - 2\overline\phi^2 |\nabla\chi_a|^2 \Delta\chi_a
      + \overline \phi^2 |\nabla\chi_a|^2\right]\\
  &\leq C(\chi)(a+1)^{N-k-1}\int_{\R^k} \left[ |\nabla\overline\phi|^2 +
\overline \phi^2\right],
\end{align*}
where the constant in the last line depends on $\chi$ but can be chosen
independent of $a$.

Combining this estimate with~\pref{eq:discrepancy} provides us with the opposite
inequality,
\begin{align}
\omega_{N-k} a^{N-k}\|\overline u - \overline v\|^2_{H^{-1}(\R^k)}
  &\leq \int_{\R^N} \chi_a^2 |\nabla\overline\phi|^2\notag\\
  &\leq \|u-v\|_{H^{-1}(\R^N)}^2 + \notag\\
  & \qquad {}+ C(\chi)(a+1)^{N-k-1}\int_{\R^k} \left[ |\nabla\overline\phi|^2 +
\overline \phi^2\right].
  \label{ineq:chi}
\end{align}
Summarising \pref{ineq:chi} and \pref{eq:discrepancy} as
\[
\|u-v\|^2_{H^{-1}(\R^N)}
  - \omega_{N-k} a^{N-k}\|\overline u-\overline v\|^2_{H^{-1}(\R^k)}
\begin{cases}
\leq  \omega_{N-k} \bigl((a+1)^{N-k}-a^{N-k}\bigr)
                         \|\overline u-\overline v\|^2_{H^{-1}(\R^k)} \\
\geq  -C(\chi)(a+1)^{N-k-1}\int_{\R^k} \left[ |\nabla\overline\phi|^2 +
\overline \phi^2\right],
\end{cases}
\]
we find the statement of the lemma.
\end{proof}

\subsection{Examples with prescribed morphology}
\label{subsec:examples}

Theorem~\ref{th:cutoff} has the following consequence: when comparing energy-per-unit-mass of structures in dimension $N$, we can include the energy-per-unit-mass of structures in dimension $k<N$, up to a correction term that decays to zero in the limit of large mass. We now use this tool to investigate the energy values of various fixed-geometry structures.

\begin{itemize}
\item A one-dimensional, lamellar, structure. The optimal energy-per-unit-mass
is $\left(\frac92\right)^{1/3}d_{uv}^{2/3}$ (Theorem~\ref{th:lower_bound_1d}, achieved in
the limit of large mass).
\item A \emph{micelle} in $N$ dimensions, i.e. a spherical particle described by
\begin{align*}
u_m(x) &:= \left\{ \begin{array}{ll} 1 & \text{if } 0<|x|< R_1,\\
0 & \text{otherwise},\end{array}\right.\\
v_m(x) &:= \left\{ \begin{array}{ll} 1 & \text{if } R_1< |x|< R_2,\\
0 & \text{otherwise}.\end{array}\right.
\end{align*}
The equal-mass criterion implies that $R_2=2^{1/N}R_1$, and by optimising with
respect to the remaining parameter $R_1$ we find that the optimal energy-per-
unit-mass is (Theorem~\ref{thm:energysphersymmono})
\[
3 \left(d_{uv} + d_{v0}\sqrt2\right)^{\frac23} \left(\log\,2 - \frac12\right)^
{\frac13} \qquad\text{for $N=2$},
\]
and
\begin{equation}
\label{num:3d-micelle}
2^{-1/3}\,3N\bigl(d_{uv}+d_{v0}2^{1-1/N}\bigr)^{2/3}
  \left(\frac{N+2-N2^{2/N}}{N(N^2-4)}\right)^{1/3}
  \qquad\text{for $N\geq 3$}.
\end{equation}
These optimal values are attained at \emph{finite} mass.

In both cases the micelle energies are larger than $\left(\frac92\right)^{1/3}d_{uv}^{2/3}
$, even when $d_{v0}=0$, implying that for large $M$ lamellar structures have lower energy per
unit mass than micelles.

\item A $k$-dimensional micelle embedded in $N$-dimensional space, similarly to
the case of lamellar structures in $N$ dimensions. A two-dimensional micelle
thus becomes a cylinderical structure in three dimensions.

The energy per unit of mass of such a structure will be lower than that of an $N$-dimensional micelle, since \pref{num:3d-micelle} is a strictly increasing function of $N$, but larger than that of a lamellar structure for large mass, by the conclusion of the previous point.

\item A monolayer in the shape of a spherical shell as in Theorem~\ref{thm:energysphersymmono}. Here the optimal energy per
unit mass can be found (in the limit of large radius $R$) by minimising~\pref
{eq:genrad} with respect to $M$:
\begin{align*}
&\left(\frac92\right)^{1/3}\left( d_{u0} + d_{uv} + d_{v0} \right)^{2/3} + (N - 1) \left
( d_{v0} - d_{u0} \right) R^{-1}  + \\
& \qquad\qquad {}+\left(\frac34\right)^{1/3}(N-1)
  \left(-\frac{3N-12}{20}(d_{u0}+d_{v0}) +\frac{3N-2}{20}d_{uv}\right) R^{-2}
    + \mathcal{O}(R^{-3}).
\end{align*}
Note that this value approaches for $R\to\infty$ the optimal one-dimensional value when $d_{u0}
=d_{v0}=0$. (Although such a choice is ruled out by~\pref{eq:ddemands}, one may
calculate the value of the energy in this case nonetheless.) In this case the
limit value is approached from above.

Alternatively, if either $d_{u0}$ or $d_{v0}$ is non-zero, then the limit value
is larger than that of the optimal lamellar structures.
\end{itemize}

Among this list, therefore, the structures with lowest energy per unit mass are the lamellar structures. It seems natural to conjecture that global minimisers also resemble cut-off lamellar structures, and have comparable energy per unit mass. On the other hand, the results of the companion paper~\cite{vanGennipPeletier07b} show that bilayer structures are unstable in a part of parameter space, and similarly Ren and Wei showed that for the pure diblock copolymer model ($u+v\equiv 1$) 'wriggled lamellar' solutions may have lower energy than straight ones~\cite{RenWei05}. Determining the morphology of large-mass global minimisers is therefore very much an open question.

\section{Discussion and conclusions}

The results discussed in this paper provide an initial  view on the properties of the energy~\pref{eq:functional}, and consequently on mixtures of block copolymers with homopolymers. The sharp-limit version of the more classical smooth-interface energy provides a useful simplification and provides us with tools that would otherwise be unavailable.

In one dimension we continued on the work of Choksi and Ren and gave a complete characterisation of the structure of one-dimensional minimisers, both on $\R$ and on a periodic cell.

In the multi-dimensional case we have proved upper and lower bounds for the energy of minimisers. These bounds both scale linearly with mass, but have different constants. The upper bound is derived from the one-dimensional minimisers, thanks to the cut-off estimate of Theorem~\ref{th:cutoff}; the results of the companion paper on the stability of mono- and bilayers \cite{vanGennipPeletier07b} suggest that for some parameter values this upper bound can be exact, while for others it is not. Similarly, the lower bound proved in Section~\ref{subsec:lower_bound} has the right scaling in terms of mass, but the constant is not sharp.

The sharpness of the estimates is especially relevant in relation to the issue of optimal morphology. A precise estimate of the energy level of energy minimisers may exclude large classes of morphologies and thus limit the possible morphology of energy minimisers.

Since we lack such a sharp estimate the question of the preferred morphology in multiple dimensions is still completely open. Part of this question is the behaviour of the morphology near the copolymer-homopolymer interface. For instance, if the preferred morphology is lamellar, does the lamellar orientation show a preference to be orthogonal, parallel, or otherwise aligned with respect to the interface? The experimental observations of for instance~\cite{KoizumiHasegawaHashimoto94,OhtaNonomura97,ZhangJinMa05} show both orthogonal and parallel alignments. Other issues are those of the penalty incurred by certain macrodomain morphologies and defects (such as chevron or loop morphologies \cite[Figures~11 and~14]{AdhikaryMichler04}).

The large-mass limit for the functional $F_1$ is equivalent to a singular-limit process at fixed mass for the functional $F_\e$~\pref{def:Fe}. As discussed in Section~\ref{subsec:partloc} the results of~\cite{PeletierRoeger06}  suggest that for certain values of the $c_i$---to be precise, for those values for which bilayer structures are stable---the functional $F_\e$ may display similar, \emph{partially localised} behaviour.

The results from this paper have some interesting physical implications. Theorem~\ref{th:cutoff} tells us that extended one-dimensional minimisers, i.e. layered structures, will have a relatively low energy (although the question whether or not these are minimisers is still open). Theorem~\ref{th:CR2} and Remark~\ref{rem:convergenceofwidth} show that these layers all have the same width. One can think of lamellar configurations like this as having all polymer molecules aligned in straight rows next to each other. Structures like 0U0, which are not to be expected on physical grounds, are a priori not forbidden in our model, but such configurations are not stationary points, as is shown in Section~\ref{subsec:connected_support}.

Depending on the surface tension coefficients very different structures can appear. As remarked in Section~\ref{subsec:d_12} the role of $d_{uv}$ is a special one. If $d_{uv}=0$, there is no repulsion between the U- and V-phases, but there is attraction, due to the $H^{-1}$-norm. Complete mixing of both phases will occur. If one of the other surface tension coefficients is zero instead, say $d_{u0}=0$, then UVU bilayers can be joined together without extra cost, and vice versa UVUVU structures can be split through the middle without increasing the energy. Physically this happens if the U- and 0-phases do not repel each other. The simplest case one can think of is if both phases consist of the same material. If both $d_{u0}>0$ and $d_{v0}>0$, it will always be energetically favourable to join different layers together, because doing so decreases the length of the energetically costly interfaces.

\appendix
\section{Spherically symmetric configurations}\label{sec:sphersym}

In this appendix we will compute the energy $F_1$ of spherically symmetric
monolayers and bilayers. In \cite{OhtaNonomura98} the energy for a spherically symmetric
bilayer in two and three dimensions is computed. We will give the energy in any
dimension $N$.

\begin{figure}[ht]
\hspace{0.1\textwidth}
\subfloat[Spherical monolayer][A spherically symmetric monolayer in two dimensions]
{
    \psfrag{A}{U}
    \psfrag{B}{V}
    \psfrag{C}{$R_0$}
    \psfrag{D}{$R_1$}
    \psfrag{E}{$R_2$}
    \includegraphics[width=0.35\textwidth]{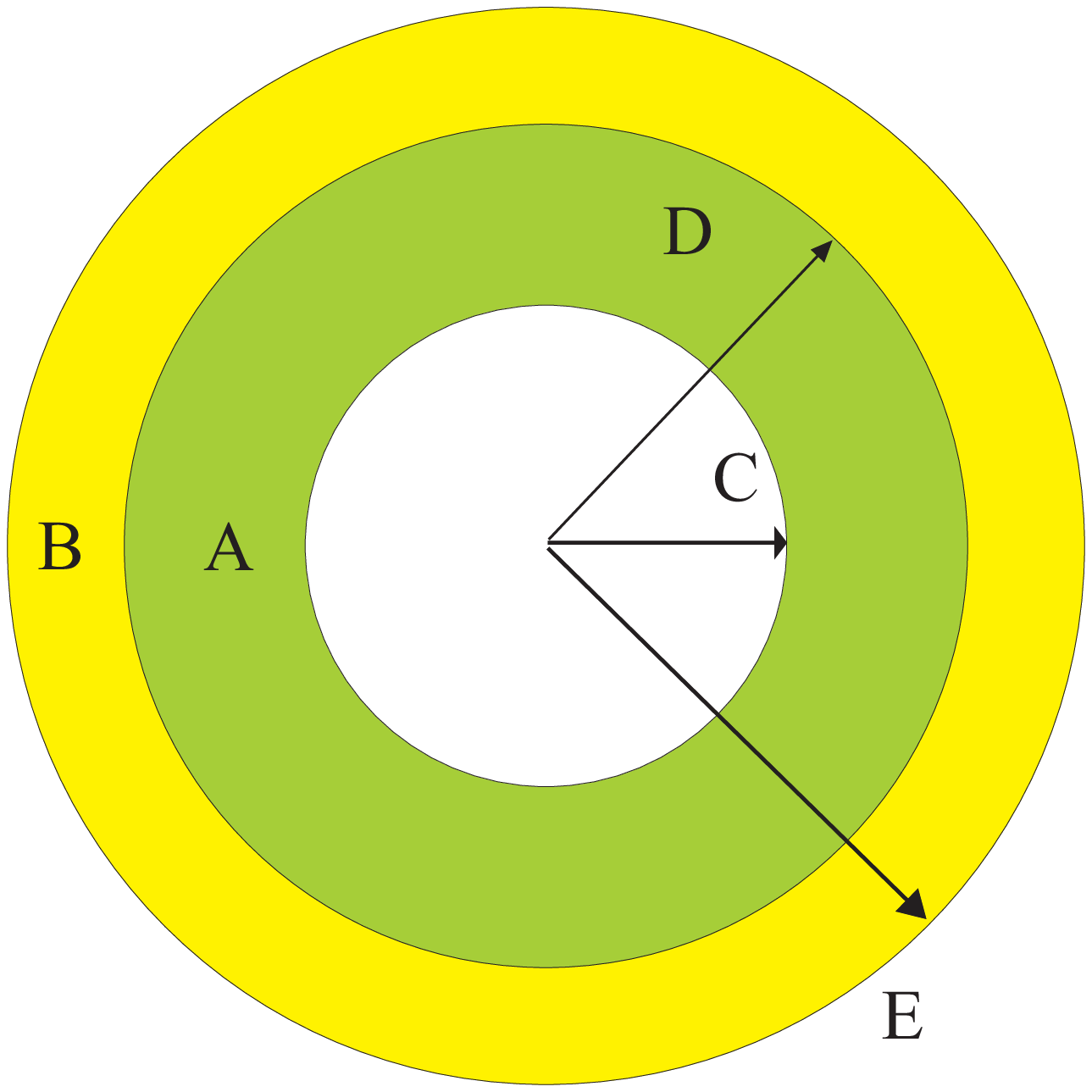}\\
    \label{fig:sphericalmonolayer}
}
\hspace{0.1\textwidth}
\subfloat[Spherical bilayer][A spherically symmetric bilayer in two dimensions]
{
    \psfrag{A}{U}
    \psfrag{B}{V}
    \psfrag{C}{$R_0$}
    \psfrag{D}{$R_1$}
    \psfrag{E}{$R_2$}
    \psfrag{F}{$R_3$}
    \includegraphics[width=0.35\textwidth]{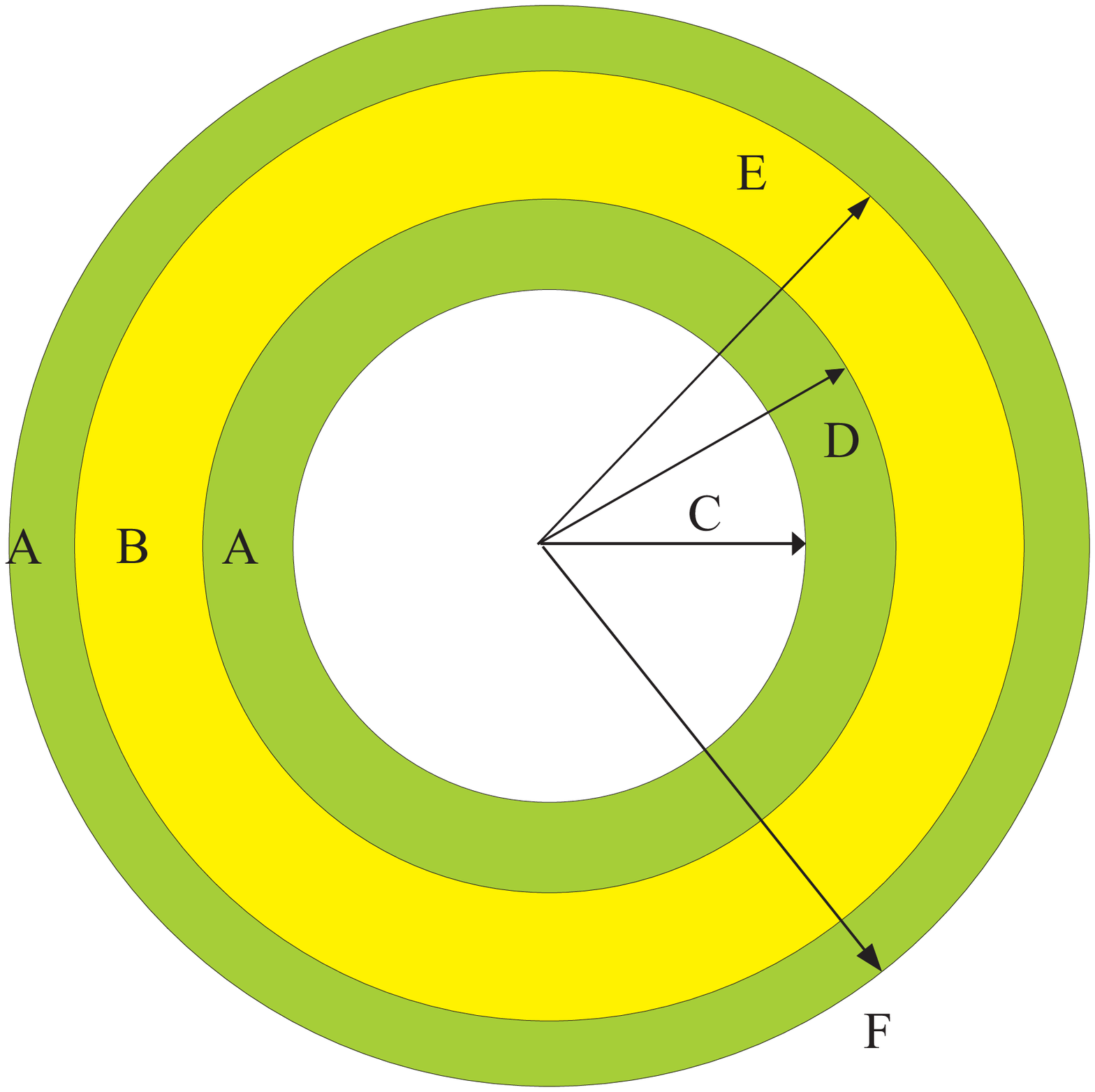}\\
    \label{fig:sphericalbilayer}
}
\caption{}
\end{figure}

A \emph{spherically symmetric monolayer with inner U-band} in $N$ dimensions consists of a spherical layer of U between radial distances $R_0$ and $R_1$ and a spherical layer of V between radial distances $R_1$ and $R_2$. An example for $N=2$ is drawn in Figure~\ref{fig:sphericalmonolayer}. Similarly, a \emph{spherically symmetric UVU bilayer} is a spherical layer of V between radial distances $R_1$ and $R_2$, flanked by two spherical layers of U, between radial distances $R_0$ and $R_1$, and $R_2$ and $R_3$ respectively. A two-dimensional example is shown in Figure~\ref{fig:sphericalbilayer}. Monolayers with inner V-band or VUV bilayers are constructed by interchanging U and V.

We will first compute the energy for monolayers. In Theorem~\ref{thm:energysphersymmono} we give the expansion in terms of the \emph{curvature} $\kappa$ of the energy per mass, for small $\kappa$. The exact expressions for $F_1$ can be found in the proof of the theorem, in (\ref{eq:sphericalmono2d}) for $N=2$ and in (\ref{eq:sphericalmonond}) for $N\geq3$.

The expansion in terms of small curvature is obtained by linearising these exact energy expressions around ``$R=
\infty$''. To this end we introduce for the monolayer the curvature $\kappa$, total U-mass $M$, and mass per (hyper-)surface area $m$:
\begin{align*}
\kappa &:= R_1^{-1},\qquad M := \omega_N (R_1^N-R_0^N),\\
m &:= \frac{M}{N \omega_N R_1^{N-1}} = \frac{M \kappa^{N-1}}{N \omega_N}.
\end{align*}
We then get
\begin{align*}
R_0 &= \kappa^{-1} \sqrt[N]{1 - N m \kappa },\\
R_2 &= \kappa^{-1} \sqrt[N]{1 + N m \kappa }.
\end{align*}

\begin{theorem}\label{thm:energysphersymmono}
Let $(u, v) \in K_1$ be a spherically symmetric monolayer with inner U-band. Fix the mass per surface area $m > 0$, then for all $N \geq 2$:
\begin{align}
\frac{F_{1}}{M}(u, v) =& \, \, m^{-1} \left( d_{u0} + d_{uv} + d_{v0} \right) +
\frac{2}{3} m^2 + (N - 1) \left( d_{v0} - d_{u0} \right) \kappa\nonumber\\
 &+ (N - 1) m \left( -\frac{1}{2} (d_{u0} + d_{v0}) + \frac{1}{15} (3 N - 2) m^3
\right) \kappa^2+ \mathcal{O}(\kappa^3),\label{eq:genrad}
\end{align}
if $\kappa \downarrow 0$.
\end{theorem}

Note that there are two configurations for a monolayer, depending on whether the
U-phase is on the inside or the outside. The theorem above states the case where
the U-phase is on the inside. The other case is found by
interchanging $d_{u0}$ and $d_{v0}$.

\smallskip

\begin{proof}[Proof of Theorem~\ref{thm:energysphersymmono}]
The proof consists of three steps. First we compute $F_1$ in terms of the radii $R_i$, then we rewrite it as a function of $\kappa, M$ and $m$. Finally the expansion is found by computing the first terms of the Taylor series  of these expressions for $\kappa\ll 1$.

The interfacial terms are computed in a straightforward manner. For the $H^{-1}$-norm we need to
compute the Poisson potential, which depends only on the radius $r$ because of
the spherical symmetry and which we will denote by $\phi(r)$. The Poisson
equation in spherical coordinates is
\begin{equation}\label{eq:poissonradial}
\left\{
\begin{array}{l}
-r^{-N+1} \left(r^{N-1} \phi'(r)\right)' = u - v \quad \text{for } r>0,\\
\phi(0) = \phi'(0) = 0.
\end{array}
\right.
\end{equation}
The solutions to this equation for $N = 2$ and $N \geq 3$ are different, and we
treat these two cases separately.
First we solve for $N = 2$ for the four different regions and we match the
solutions under the condition that $\phi \in C^1(\R^N)$. This gives
\[
\phi(r) = \left\{ \begin{array}{ll}
0
  &\mbox{$r \in (0, R_0)$,}\\
-\frac{1}{4}r^2 + \frac{1}{2} R_0^2 \log r + \frac{1}{4} R_0^2 - \frac{1}{2}
R_0^2 \log R_0
  &\mbox{$r \in (R_0, R_1)$,}\\
\frac{1}{4}r^2 + \frac{1}{2} \left(R_0^2 - 2 R_1^2\right) \log r + R_1^2 \log
R_1 - \frac{1}{2} R_0^2 \log R_0 - \frac{1}{2} R_1^2 + \frac{1}{4} R_0^2
  &\mbox{$r \in (R_1, R_2)$,}\\
\frac{1}{4} \left( 2 R_1^2 - R_0^2 \right) \left( 1 - \log (2 R_1^2 - R_0^2)
\right) + R_1^2 \log R_1 - \frac{1}{2} R^2 \log R_0 - \frac{1}{2}R_1^2 + \frac
{1}{4}R_0^2
  &\mbox{$r>R_2$.}  \end{array} \right.
\]
Note that $\phi$  is constant on $[R_2, \infty)$: the solution can not have a
term proportional to $\log r$ on this interval, since $\phi \in W^{1,2}(\R^N)$
and
\[
\int_{R_2}^{\infty} |\partial_r \log r|^2 r \, dr = \int_{R_2}^{\infty} \left( \frac
{1}{r} \right)^2 r \, dr = \infty.
\]
This means that $\phi'(R_0) = \phi'(R_2) = 0$. We now compute the norm via
\[
\|u - v\|_{H^{-1}(\Omega)}^2 = 2 \pi \left(\int_{R_0}^{R_1} \phi(r) r \, dr -
\int_{R_1}^{R_2} \phi(r) r \, dr\right).
\]
For  $N = 2$, we then find
\begin{align}
\frac{1}{2 \pi} F_1(u, v) &= R_0 d_{u0} + R_1 d_{uv} + R_2^2 d_{v0}\nonumber\\
&\hspace{0.4cm} - \frac{1}{4} R_1^4 + \frac{1}{4} R_0^2 R_1^2 - \frac{1}{4}
R_0^4 \log  R_0 - R_1^2 \left( R_1^2 - R_0^2 \right) \log R_1 \nonumber\\
&\hspace{1.5cm} + \frac{1}{8} \left( 2 R_1^2 - R_0^2 \right)^2 \log\left( 2
R_1^2 - R_0^2 \right),\label{eq:sphericalmono2d}
\end{align}
where the radii are related by $R_2^2 - R_1^2 = R_1^2 - R_0^2$.

Analogously solving for $N \geq 3$ we find
\[
\phi(r) = \left\{ \begin{array}{ll} 0
  &\mbox{ if $r \in (0, R_0)$,}\\
\frac{-1}{N (N - 2)} R_0^N r^{-N+2} - \frac{1}{2 N} r^2 + \frac{1}{2 (N - 2)}
R_0^2
  &\mbox{ if $r \in (R_0, R_1)$,}\\
\frac{-1}{N (N - 2)} \left( R_0^N - 2 R_1^N \right) r^{-N + 2} + \frac{1}{2 N}
r^2 + \frac{1}{2 (N - 2)} \left( R_0^2 - 2 R_1^2 \right)
  &\mbox{ if $r \in (R_1, R_2)$,}\\
\frac{1}{2 (N - 2)} \left( \left( 2 R_1^N - R_0^N \right)^{\frac{2}{N}} - 2
R_1^2 + R_0^2 \right)
  &\mbox{ if $r >R_2$,}\end{array} \right.
\]
and compute the norm. This leads to
\begin{align}
\frac{F_1}{N \omega_N}(u,v) &= R_0^{N-1} d_{u0} + R_1^{N-1} d_{uv} + R_2^{N-1}
d_{v0}\nonumber\\
&\hspace{0.4cm}+ \frac{1}{N^2 - 4} \left( R_0^{N+2} - R_2^{N+2} \right) + \frac
{2}{N (N-2)} R_1^2 \left( R_1^N - R_0^N \right),\label{eq:sphericalmonond}
\end{align}
where the radii are related by $R_2^N - R_1^N = R_1^N - R_0^N$.

Rewriting our results in terms of $\kappa, M$ and $m$ gives, for $N=2$,
\begin{align}
\frac{F_1}{M}(u, v) =& \, M^{-1} \left(\sqrt{1 - 2 m \kappa } d_{u0} + d_{uv} +
\sqrt{1 + 2 m \kappa } d_{v0} \right) - \frac{1}{2} \kappa^{-2} \nonumber\\
 &+ \frac12 \left( \frac{1}{4} m^{-1} \kappa^{-3}+ \kappa^{-2}  + m \kappa^{-1}
\right) \log(1 + 2 m \kappa )\nonumber\\
 &- \frac12 \left( \frac{1}{4} m^{-1} \kappa^{-3} - \kappa^{-2} + m \kappa^{-1}
\right) \log(1 - 2 m \kappa ).\nonumber
\end{align}
For the monolayer with $N\geq3$ we get
\begin{align*}
\frac{F_1}{M}(u, v) =& \, m^{-1} \left( (1 - N m \kappa)^{\frac{N-1}{N}} d_{u0}
+ d_{uv} + (1 + N m \kappa)^{\frac{N-1}{N}} d_{v0} \right)\\
&+ \frac{1}{N^2 - 4} m^{-1} \left( (1 - N m \kappa)^{\frac{N+2}{N}} - (1 + N m
\kappa)^{\frac{N+2}{N}} \right) \kappa^{-3}  + \frac{2}{N - 2} \kappa^{-2}.
\end{align*}
We now expand in terms of $\kappa \ll 1$ to get the result.
\end{proof}

Next we turn to the bilayer. Here we follow the same route as before. Theorem~\ref{thm:energysphersymbilay} states the expansion in small curvature; the exact expressions for $F_1$ can be found in the proof in (\ref{eq:sphericalbi2d}) for $N=2$ and in (\ref{eq:sphericalbind}) for $N\geq 3$. Define $R > 0$ via
$R^N = \frac12 \left( R_1^N + R_2^N \right)$, then for the bilayer we introduce curvature $\kappa$, total U-mass $M$ and mass per (hyper-)surface area $m$ for $N\geq 1$ as follows:
\begin{align*}
\kappa &:= R^{-1},\qquad M:=\omega_N(R_2^N-R_1^N),\\
m &:= \frac{M \kappa^{N-1}}{N \omega_N}.
\end{align*}
Then
\begin{align*}
R_0 &= \kappa^{-1} \sqrt[N]{\left(1 - N m \kappa\right)}, \quad R_1 = \kappa^
{-1} \sqrt[N]{\left(1 - \frac12 N m \kappa\right)},\\
R_2 &= \kappa^{-1} \sqrt[N]{\left(1 + \frac12 N m \kappa\right)}, \quad R_3 =
\kappa^{-1} \sqrt[N]{\left(1 + N m \kappa\right)}.
\end{align*}
Note that here we have chosen the radii such that the inner and outer U-band have equal mass: $R_3^2 - R_2^2 = R_1^2 - R_0^2$. This is not the optimal choice, in the sense that a small change in the relative thicknesses of the inner and outer monolayers might improve the energy slightly. We expect this to be a small effect, however. 
\begin{theorem}\label{thm:energysphersymbilay}
Let $(u, v) \in K_1$ be a spherically symmetric UVU bilayer. Fix the mass per surface area $m > 0$, then for all $N \geq 2$:
\begin{align}
\frac{F_{1}(u, v)}{M} &= 2 m^{-1} (d_{u0} + d_{uv}) + \frac16 m^2\nonumber\\
&\hspace{0.4cm}+ (N - 1) m \left( -\left( d_{u0} + \frac14 d_{uv} \right) +
\frac{11}{240} (3 N - 2) m^3 \right)  \kappa^2 + \mathcal{O}(\kappa^4),\label{eq:bilayerexpansion}
\end{align}
if $\kappa \downarrow 0$.
\end{theorem}

An analogous result and proof corresponding to the VUV bilayer is constructed by replacing $d_{u0}$ by $d_{v0}$.

\begin{proof}[Proof of Theorem~\ref{thm:energysphersymbilay}]
As in the proof of theorem \ref{thm:energysphersymmono} we follow three steps. First we compute $F_1$ in terms of the radii $R_i$. The resulting expression we rewrite in terms of $\kappa, M$ and $m$ and finally we find the expansion in terms of $\kappa \ll 1$ by calculating the first terms of a Taylor series.

The main problem in the first step consists of deriving the Poisson potential that solves (\ref{eq:poissonradial}).
For $N = 2$ we find
\[
\phi(r) =
\left\{ \begin{array}{ll}
0 &\mbox{ if $r \in (0, R_0)$,}\vspace{0.1cm}\\
-\frac14 r^2 + \frac12 R_0^2 \log r + \frac14 R_0^2 - \frac12 R_0^2 \log R_0 &
\mbox{ if $r \in (R_0, R_1)$,}\vspace{0.1cm}\\
\frac14 r^2 + \frac12 (R_0^2 - 2 R_1^2) \log r + R_1^2 \log R_1&\\
\hspace{0.4cm} - \frac12 R_0^2 \log R_0 - \frac12 R_1^2 + \frac14 R_0^2 &\mbox{
if $r \in (R_1, R_2)$,}\vspace{0.1cm}\\
-\frac14 r^2 + \left(\frac12 R_0^2 - R_1^2 + R_2^2\right) \log r - \frac12 R_0^2
\log R_0&\\
\hspace{0.4cm} + R_1^2 \log R_1 - R_2^2 \log R_2 + \frac14 R_0^2 - \frac12 R_1^2
+ \frac12 R_2^2 &\mbox{ if $r \in (R_2, R_3)$,}\vspace{0.1cm}\\
-\frac14 R_3^2 + \left(\frac12 R_0^2 - R_1^2 + R_2^2\right) \log R_3 - \frac12
R_0^2 \log R_0&\\
\hspace{0.4cm} + R_1^2 \log R_1 - R_2^2 \log R_2 + \frac14 R_0^2 - \frac12 R_1^2
+ \frac12 R_2^2 &\mbox{ if $r >R_3$.}
\end{array} \right.
\]

For $N \geq 3$ we have
\[
\phi(r) =
\left\{ \begin{array}{ll}
0 &\mbox{ if $r \in (0, R_0)$,}\\
-\frac1{N(N-2)} R_0^N r^{-N+2} - \frac1{2N} r^2 + \frac1{2(N-2)} R_0^2 &\mbox
{ if $r \in (R_0, R_1)$,}\\
-\frac1{N(N-2)} \left(R_0^N - 2 R_1^N\right) r^{-N+2} + \frac1{2N} r^2 + \frac1
{2(N-2)} \left(R_0^2 - 2 R_1^2\right) &\mbox{ if $r \in (R_1, R_2)$,}\\
-\frac1{N(N-2)} \left(R_0^N - 2 R_1^N + 2 R_2^N\right) r^{-N+2} - \frac1{2N} r^2
+ \frac1{2(N+2)} \left(R_0^2 - 2 R_1^2 + 2 R_2^2\right) &\mbox{ if $r \in (R_2,
R_3)$,}\\
\frac1{2(N-2)} \left(R_0^2 - 2 R_1^2 + 2 R_2^2 - R_3^2\right) &\mbox{ if $r>R_3
$.}
\end{array} \right.
\]

We then proceed in the same way as for Theorem \ref{thm:energysphersymmono} to find, for $N = 2$,
\begin{multline}\label{eq:sphericalbi2d}
\frac{1}{2 \pi} F_1(u, v) = (R_0 + R_3) d_{u0} + (R_1 + R_2) d_{uv}+  \frac1{16}
\left(R_0^4 - R_3^4\right) +\\
\begin{aligned}
&\hspace{1.5cm} + \left( \frac12 R_0^2 R_1^2 + \frac12 R_0^2 R_2^2 - \frac14
R_0^2 R_3^2 \right) \log R_0 + \left(\frac12 R_0^2 R_1^2 - R_1^2 R_2^2 + \frac12
R_1^2 R_3^2\right) \log R_1 +\\
&\hspace{1.5cm} + \left(-\frac12 R_0^2 R_2^2 + R_1^2 R_2^2 - \frac12 R_2^2 R_3^2
\right) \log R_2 + \left(\frac14 R_0^2 R_3^2 - \frac12 R_1^2 R_3^2 + \frac12
R_2^2 R_3^2\right) \log R_3,
\end{aligned}
\end{multline}
where $R_3^2 - R_2^2 = \frac12 (R_2^2 - R_1^2) = R_1^2 - R_0^2$.

For a bilayer with $N \geq 3$ we have
\begin{multline}\label{eq:sphericalbind}
\frac{1}{N \omega_N} F_1(u, v) = (R_0^{N-1} + R_3^{N-1}) d_{u0} + (R_1^{N-1} +
R_2^{N-1}) d_{uv}+ \frac1{2 N (N+2)} \left(R_0^{N+2} - R_3^{N+2} \right)\\
{} + \frac1{2 N (N-2)} \left( -4 R_0^N R_1^2 + 4 R_0^N R_2^2 - 8 R_1^N R_2^2 +
R_0^{N+2} + 4 R_1^{N+2} + 4 R_2^{N+2} - R_3^{N+2} \right),
\end{multline}
where $R_3^N - R_2^N = \frac12 (R_2^N - R_1^N) = R_1^N - R_0^N$.

Rewriting these results in terms of $\kappa, M$ and $m$ gives, for $N = 2$
\begin{align*}
\frac{F_{1}(u, v)}{M} &= m^{-1} \left[ d_{u0} \left( (1 - 2 m \kappa)^{1/2} + (1
+ 2 m \kappa)^{1/2} \right) + d_{uv} \left( (1 - m \kappa)^{1/2} + (1 + m
\kappa)^{1/2} \right) \right]\\
&\hspace{1.4cm}- \frac12 \kappa^{-2}  - \frac12 \left( \frac14 m^{-1} \kappa^
{-3}  - \kappa^{-2}  + m \kappa^{-1}  \right) \log(1 - 2 m \kappa)\\
&\hspace{1.4cm} + \frac12 \left( \frac14 m^{-1} \kappa^{-3}  + \kappa^{-2}  + m
\kappa^{-1} \right) \log(1 + 2 m \kappa)\\
&\hspace{1.4cm} - \frac12 \left( \kappa^{-2}  + m \kappa^{-1}  \right) \log(1 +
m \kappa) + \frac12 \left( -\kappa^{-2}  + m \kappa^{-1}  \right) \log(1 - m
\kappa).
\end{align*}

For $N \geq 3$ we have
\begin{align*}
\frac{F_{1}(u, v)}{M} &= m^{-1} \left[\left( (1 - N m \kappa)^{\frac{N-1}{N}} +
(1 + N m \kappa)^{\frac{N-1}{N}} \right) d_{u0}\right.\\
&\hspace{1.4cm} \left.  + \left( \left(1 - \frac12 N m \kappa\right)^{\frac{N-1}
{N}} + \left(1 + \frac12 N m \kappa\right)^{\frac{N-1}{N}} \right) d_{uv}\right]
\\
&\hspace{0.4cm}+ \frac1{N^2 - 4} m^{-1} \left( (1 - N m \kappa)^{\frac{N+2}{N}}
- (1 + N m \kappa)^{\frac{N+2}{N}} \right) \kappa^{-3}\\
&\hspace{0.4cm}+ \frac{1}{N-2} \left( \left(1 - \frac12 N m \kappa\right)^
{\frac2N} + \left(1 + \frac12 N m \kappa\right)^{\frac2N} \right) \kappa^{-2}.
\end{align*}
The result (\ref{eq:bilayerexpansion}) now follows from expanding in $\kappa \ll 1$.
\end{proof}

\bibliographystyle{alpha}
\bibliography{bib_revision}

\end{document}